\documentclass[iop]{emulateapj}
\usepackage{epsfig}
\usepackage{natbib}
\usepackage{amssymb}
\usepackage{amsbsy}
\usepackage{natbib}
\usepackage{subfigure}
\usepackage[mathcal]{euscript}
\usepackage{float}
\usepackage{amsmath}
\usepackage{tabularx}

\bibpunct{(}{)}{;}{a}{}{,}

\begin{document}
\title{The Correlation Between Mixing Length and Metallicity on the Giant Branch: Implications for Ages in the \textit{Gaia} Era}
\author{
Jamie Tayar\altaffilmark{1}, 
Garrett Somers\altaffilmark{2,1}, 
Marc H.~Pinsonneault\altaffilmark{1}, 
Dennis Stello\altaffilmark{3,4,5}, 
Alexey Mints\altaffilmark{6,4}, 
Jennifer A.~Johnson\altaffilmark{1}, 
O.~Zamora\altaffilmark{7,8}, 
D.~A.~Garc{\'i}a-Hern{\'a}ndez\altaffilmark{7,8}, 
Claudia Maraston\altaffilmark{9}, 
Aldo Serenelli\altaffilmark{10}, 
Carlos Allende Prieto\altaffilmark{11}, 
Fabienne A.~Bastien\altaffilmark{12,13}, 
Sarbani Basu\altaffilmark{14}, 
J.~C.~Bird\altaffilmark{2}, 
R.~E.~Cohen\altaffilmark{15}, 
Katia Cunha\altaffilmark{16}, 
Yvonne Elsworth\altaffilmark{17}, 
Rafael A.~Garc{\'i}a\altaffilmark{18}, 
Leo Girardi\altaffilmark{19}, 
Saskia Hekker\altaffilmark{6,4}, 
Jon Holtzman\altaffilmark{20}, 
Daniel Huber\altaffilmark{3,21,4}, 
Savita Mathur\altaffilmark{22}, 
Szabolcs M{\'e}sz{\'a}ros\altaffilmark{23,24}, 
B.~Mosser\altaffilmark{25}, 
Matthew Shetrone\altaffilmark{26}, 
Victor Silva Aguirre\altaffilmark{4}, 
Keivan Stassun\altaffilmark{2}, 
Guy S.~Stringfellow\altaffilmark{27}, 
Gail Zasowski\altaffilmark{28}, and 
A.~Roman-Lopes\altaffilmark{29} 
}
\altaffiltext{1}{Department of Astronomy, Ohio State University, 140 W 18th Ave, OH 43210, USA}
\altaffiltext{2}{Department of Physics and Astronomy, Vanderbilt University, 6301 Stevenson Circle, Nashville, TN, 37235}
\altaffiltext{3}{Sydney Institute for Astronomy (SIfA), School of Physics, University of Sydney, NSW 2006, Australia}
\altaffiltext{4}{Stellar Astrophysics Centre, Department of Physics and Astronomy, Aarhus University, Ny Munkegade 120, DK-8000 Aarhus C, Denmark}
\altaffiltext{5}{School of Physics, University of New South Wales, NSW 2052, Australia}
\altaffiltext{6}{Max-Planck-Institut f\"ur Sonnensystemforschung, Justus-von-Liebig-Weg 3, 37077 G\"ottingen, Germany}
\altaffiltext{7}{Instituto de Astrofísica de Canarias (IAC), Vía Lactea s/n, E-38205 La Laguna, Tenerife, Spain}
\altaffiltext{8}{Departamento de Astrofísica, Universidad de La Laguna (ULL), E-38206 La Laguna, Tenerife, Spain}
\altaffiltext{9}{ICG - University of Portsmouth, Burnaby Road, PO1 3FX, Portsmouth}
\altaffiltext{10}{Institute of Space Sciences (CSIC-IEEC), Carrer de Can Magrans, Barcelona, 08193, Spain}
\altaffiltext{11}{Instituto de Astrofisica de Canarias, 38205 La Laguna, Tenerife, Spain}
\altaffiltext{12}{Department of Astronomy and Astrophysics, 525 Davey Lab, The Pennsylvania State University, University Park, PA 16803}
\altaffiltext{13}{Hubble Fellow}
\altaffiltext{14}{Department of Astronomy, Yale University, New Haven, CT 06511, USA}
\altaffiltext{15}{Departamento de Astronom\'{i}a, Universidad de Concepci\'{o}n, Casilla 160-C, Concepci\'{o}n, Chile}
\altaffiltext{16}{Observat\'{o}rio Nacional - MCTI, Brazil}
\altaffiltext{17}{School of Physics and Astronomy, University of Birmingham, Birmingham B15 2TT, UK}
\altaffiltext{18}{Laboratoire AIM, CEA/DRF-CNRS, Universit\'e Paris 7 Diderot, IRFU/SAp, Centre de Saclay, 91191, Gif-sur-Yvette, France}
\altaffiltext{19}{Osservatorio Astronomico di Padova-INAF, Vicolo dell’Osservatorio 5, I-35122 Padova, Italy}
\altaffiltext{20}{New Mexico State University, Las Cruces, NM 88003, USA}
\altaffiltext{21}{SETI Institute, 189 Bernardo Avenue, Mountain View, CA 94043, USA}
\altaffiltext{22}{Space Science Institute, 4750 Walnut Street Suite 205, Boulder, CO 80301, USA}
\altaffiltext{23}{ELTE Gothard Astrophysical Observatory, H-9704 Szombathely, Szent Imre herceg st. 112, Hungary}
\altaffiltext{24}{Premium Postdoctoral Fellow of the Hungarian Academy of Sciences}
\altaffiltext{25}{LESIA, Observatoire de Paris, PSL Research University, CNRS, Universit\'e Pierre et Marie Curie, Universit\'e Paris Diderot, 92195 Meudon, France}
\altaffiltext{26}{University of Texas at Austin, McDonald Observatory, 32 Fowlkes Rd, McDonald Observatory, Tx 79734-3005}
\altaffiltext{27}{Center for Astrophysics and Space Astronomy, University of Colorado, 389 UCB, Boulder, Colorado, 80309-0389}
\altaffiltext{28}{Department of Physics and Astronomy, JHU, 3400 N. Charles St. Baltimore, MD 21218}
\altaffiltext{29}{Departamento de Fısica, Facultad de Ciencias, Universidad de La Serena, Cisternas 1200, La Serena, Chile}

\begin{abstract}
In the updated APOGEE-\textit{Kepler} catalog, we have asteroseismic and spectroscopic data for over 3000 first ascent red giants. Given the size and accuracy of this sample, these data offer an unprecedented test of the accuracy of stellar models on the post-main-sequence. When we compare these data to theoretical predictions, we find a metallicity dependent temperature offset with a slope of around 100 K per dex in metallicity. We find that this effect is present in all model grids tested and that theoretical uncertainties in the models, correlated spectroscopic errors, and shifts in the asteroseismic mass scale are insufficient to explain this effect.
Stellar models can be brought into agreement with the data if a metallicity dependent convective mixing length is used, with $ \Delta\alpha_{\rm ML, YREC} \sim 0.2$ per dex in metallicity, a trend inconsistent with the predictions of three dimensional stellar convection simulations. If this effect is not taken into account, isochrone ages for red giants from the \textit{Gaia} data will be off by as much as a factor of 2 even at modest deviations from solar metallicity ([Fe/H]=$-$0.5). 
\end{abstract}

\keywords{stars: evolution}

\section{Introduction}
\setcounter{footnote}{0}
The theory of stellar structure and evolution makes a rich web of predictions about the life histories of stars.  Many predictions of this theory have proven accurate, and in particular the agreement between the predicted and actual positions of the core-hydrogen burning main sequence in the Hertzsprung-Russell diagram was a major triumph for 20th century astrophysics.  The situation for more evolved stars, however, has been more challenging to evaluate.  The temperature locus of evolved red giants is sensitive to the input physics in general and to the efficiency of stellar convection in particular. This efficiency of stellar convection, an inherently three dimensional process, has typically been parameterized in one dimensional stellar models as an effective mixing length \citep{Bohm-Vitense1958}. Testing the validity of models in this more evolved regime has historically been difficult due to the lack of both fundamental masses and reliable absolute spectroscopic measurements of temperature and detailed abundances for large samples of stars, although small samples have generally indicated reasonable agreement or only small discrepancies \citep[e.g.][]{Takeda2016, Huber2012}. 

Correct modeling of the position of the red giant branch on the Hertzsprung-Russell diagram is now of increased importance in light of the recently released {\it Gaia} data  \citep{Gaia1}. {\it Gaia}'s ability to provide accurate luminosities for many red giants, covering a much wider metallicity and age range than {\it Hipparcos} could, means that giants could be used to map galactic formation and evolution if we were able to correctly interpret their ages. Because of the steepness of the red giant branch in the HR diagram, and the relatively small effect of age, stars with a given luminosity and temperature can have widely different derived ages, depending on the adopted mixing length \citep{FreytagSalaris1999}.

In this paper we test the predictions of commonly used stellar models against spectroscopic and asteroseismic data from the APOKASC survey (Pinsonneault et al. 2016, in prep). Our test of red giant branch stellar evolution begins with the input physics for the theoretical models.  There is a consensus set of physics used in standard stellar models (see Section 2), which leads to robust model predictions for solar-type stars on the main sequence.  There are numerous sources of uncertainty in stellar modeling due to the extensive number of physical inputs, including the  nuclear reaction rates, convective overshoot, opacities, equation of state, diffusion, and the outer boundary condition. The single largest uncertainty in theoretical predictions about the locus of the red giant branch is the convective efficiency, or mixing length $\alpha$, which is typically calibrated to reproduce the known solar radius and luminosity at the solar age \citep[e.g.][]{Bressan2012, Choi2016}.  There is, however, no strong a priori reason why such a calibration should apply to evolved stars.  In fact, three dimensional convection simulations have indicated that the mixing length should depend on parameters like luminosity, gravity, and metallicity \citep{Trampedach2014, Magic2015}. There has also been recent observational evidence that a solar mixing length is not the best fit to every star \citep[e.g.][]{Bonaca2012, Metcalfe2014, Mann2015, Saio2015b, Wu2015}. 
For the purposes of this particular study we treat the mixing length as a free parameter that we use to quantify the metallicity dependent offset in temperature between theory and data.

\section{Methods}
 The first ingredient is a comprehensive spectroscopic data set, here provided by the APOGEE survey \citep{Majewski2015, Holtzman2015}.  This survey employed an automated pipeline analysis \citep{Nidever2015, GarciaPerez2015} of high resolution H-band spectra obtained for more than 100,000 stars \citep{Zasowski2013} collected using the 2.5 meter Sloan Digital Sky Survey telescope \citep{Gunn2006}. Crucially, this survey included over 6,000 red giants with asteroseismic data from the \textit{Kepler} mission.  For these stars, we took the measured frequency of maximum power ($
\nu_{\rm max}$) and large frequency spacing ($\Delta\nu$) and used the seismic scaling relations \citep[e.g.][]{KjeldsenBedding1995} to compute a mass and surface gravity (see Section 3.3 for a discussion of proposed alterations to the scaling relations). $$\frac{M}{M_\sun}\propto\left(\frac{\nu_{\rm max}}{\nu_{\rm max, \sun}}\right)^3\left(\frac{\Delta\nu}{\Delta\nu_\sun}\right)^{-4}\left(\frac{T_{\rm eff}}{T_{\rm eff, \sun}}\right)^{3/2}$$
$$\frac{g}{g_\sun}\propto\left(\frac{\nu_{\rm max}}{\nu_{\rm max, \sun}}\right)\left(\frac{T_{\rm eff}}{T_{\rm eff, \sun}}\right)^{1/2} $$
This allowed us to compare the data to theoretical tracks in the theoretical variables of mass, temperature, and gravity rather than observational spaces like color or magnitude (see Section 3).

 For this analysis, we used the second version of the combined APOGEE-\textit{Kepler} data set (APOKASC, Pinsonneault et al. 2016, in prep). This is an update of the \citet{Pinsonneault2014} APOKASC catalog containing thousands more stars, improved seismic analysis using the full \textit{Kepler} data set from a larger number of analysis pipelines (Elsworth et al. 2016, in prep), and improved spectroscopic parameters from Data Release 13  (DR13, Holtzman et al, in prep.) of the Sloan Digital Sky Survey IV (Blanton et al., in prep.). Of the 11876 APOGEE stars with {\it Kepler} data, 8887 were not flagged as dwarfs and were checked for solar-like oscillations. Of those stars, 6076 were returned as giants whose seismic frequency pattern reliably indicated whether they were core helium burning (clump, secondary clump) or not (red giant branch, asymptotic giant branch). For our analysis, we use only stars that the seismic analysis indicated are first ascent red giants in order to avoid uncertainties in mass loss, and therefore initial stellar mass, that arise when stars pass through the tip of the giant branch to the red clump.  We used stars with asteroseismic scaling relation masses and surface gravities computed using the $\nu_{\rm max}$ and $\Delta\nu$ returned by the OCT pipeline \citep{Hekker2010} based on the Kepler Asteroseismic Science Operations Center \citep[KASOC,][]{HandbergLund2014}  time series with the appropriate solar values for that pipeline ($\Delta\nu_\sun = 135.045 $, $\nu_{{\rm max}, \sun} = 3139$, T$_{\rm eff, \sun} = 5771.8 $). However, in Section 3, we confirm that using a different seismic analysis pipeline and various proposed corrections to the scaling relations do not significantly affect our results. 

For the following discussion, we restricted the sample to the 3217 stars with OCT scaling relation masses between 0.6 and 2.6 M$_\sun$, and values available for each seismic and spectroscopic parameter.  Our mass limits were chosen to exclude approximately the highest and lowest 0.1 \% of stars in the APOKASC sample, where different analysis pipelines are most likely to disagree about the stellar properties or fail to report values. We note that the stars in our sample have log(g) values from 1.1 to 3.3, although most of the stars in our sample are on the lower giant branch. The data table and model grid used in this analysis are available online.\footnote{ www.astronomy.ohio-state.edu/$\sim$tayar/MixingLength.htm}

\begin{figure}
\begin{center}
\includegraphics[width=9cm]{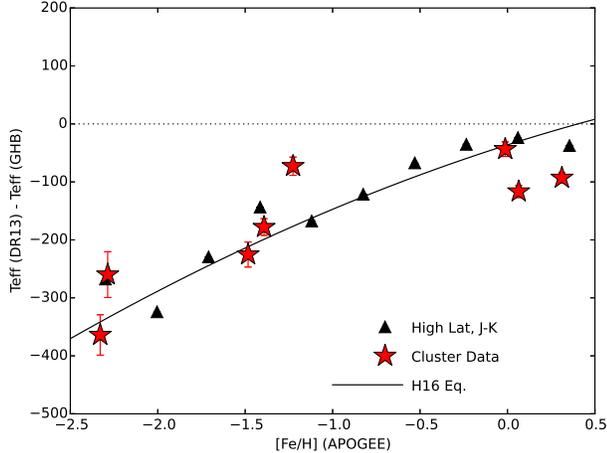}
\caption{The difference between the uncorrected Data Release 13 APOGEE temperatures and photometric temperatures computed using the J-K color temperature relations of \citet{GonzalezHernandezBonifacio2009} for clusters (red stars) and giants in low extinction fields binned by metallicity (black triangles). The temperature correction fit by Holtzman et al. (2016, in prep) is shown as a black solid line. Note that the [Fe/H] used here represents the bulk metallicity rather than the specific iron abundance.}
\label{Fig:teffcor}
\end{center}
\end{figure}

Of particular relevance to this paper, post-release analysis of DR13 spectroscopic data found that APOGEE temperatures required a metallicity dependent correction to match the photometric temperature scale (details available online\footnote{www.sdss.org/dr13/irspec/parameters}, documentation to be published in Holtzman et al. 2016, in prep). A correction was determined using the \citet{GonzalezHernandezBonifacio2009} J-K color temperature relation in low extinction fields (see Figure \ref{Fig:teffcor}); we note that this correction is consistent with or larger than the correction that would have been assumed using clusters, the \citet{RamirezMelendez2005} V-K temperature relation, or angular diameter temperatures collected in \citet{RamirezMelendez2005}. We refer the interested reader to Holtzman et al. (2016, in prep) for a more detailed discussion, but assert that the temperature offsets reported in this paper represent a lower limit on the metallicity dependence of the difference between the data and theoretical predictions.

There have also been suggestions that the APOGEE metallicity scale needs a correction to match the literature metallicities quoted in \citet{Holtzman2015}. Specifically, the APOGEE metallicities of open clusters are on average about 0.07 dex more metal poor than literature values, while the globular clusters are on average about 0.13 dex more metal rich. However, given the relatively small number of clusters, it was somewhat difficult to determine the most appropriate form of the correction. If we assume that it is linear, the best fit is $\Delta$[Fe/H]$=-0.135$[Fe/H]-0.050. Because of our uncertainty on the form of this correction, we chose not to apply it in this paper. However, on a few key plots, we indicate what effect the metallicity correction would have, and show that in all cases it makes the effect we report here stronger and more significant.

\begin{table}[htbp]
\caption{Summary of the input physics used in our main YREC and PARSEC model grids.}
\begin{tabularx} {.48\textwidth}{>{\raggedright\arraybackslash}X|>{\raggedright\arraybackslash}X|>{\raggedright\arraybackslash}X}  
\hline\hline
Parameter & YREC & PARSEC \\ \hline
Atmosphere & Gray & Gray \\ 
$\alpha$-enhancement & Yes & No \\ 
Convetive Overshoot & No & Yes \\ 
Diffusion & No & Yes \\ 
Equation of State & OPAL+SCV & FREEEOS \\ 
High Temperature Opacities& OPAL & OPAL \\ 
Low Temperature Opacities & \citet{Ferguson2005} & {\AE}SOPUS \citep{MarigoAringer2009} \\ 
Mixing Length & 1.22, 1.72, 2.22 & {1.74} \\ 
Mixture & \citet{GrevesseSauval1998} & \citet{GrevesseSauval1998}+\citet{Caffau2011} \\ 
Nuclear Reaction Rates & \citet{Adelberger2011} & \citet{Bressan2012} \\ 
Weak Screening & \citet{Salpeter1954} & \citet{Dewitt1973}+\citet{Graboske1973} \\ 
BBN He & 0.2485 & 0.2485 \\ 
Solar X & 0.709306 & {0.7091328} \\ 
Solar Y & 0.272683 & {0.2756272} \\ 
Solar Z &{0.018011} & {0.01524} \\ 
$[$Fe/H$]$ Range & -2.0 to +0.6 & -2.2 to +0.5 \\
$[\alpha$/Fe$]$ Values & 0.0, +0.2, +0.4 & 0.0\\
Mass Range & 0.6 M$_\sun$ to 2.6 M$_\sun$ & 0.1 M$_\sun$ to 20 M$_\sun$ \\\hline
\end{tabularx}
\label{Table:physics}
\end{table}

These data were compared to two different grids of stellar evolution models (see Table \ref{Table:physics} for a summary). The first was a grid run using the Yale Rotating Evolution Code \citep[YREC,][with updates as discussed in \citealt{vanSadersPinsonneault2012}] {Pinsonneault1989}. We ran a grid of masses (0.6 M$_\sun$ to 2.6 M$_\sun$ in 0.1 M$_\sun$ increments), metallicities ([Fe/H]=$-2.0$ to $+0.6$ in steps of 0.2), and $\alpha$-element enhancements (+0.0, +0.2, and +0.4) to generate tracks. These models use a \citet{GrevesseSauval1998} elemental mixture, with the updated OPAL equation of state \citep{Rogers1996, RogersNayfonov2002}, OPAL opacity tables \citep{IglesiasRogers1996} and gray atmospheres. Alpha enhanced models use alpha enhanced starting models and opacity tables, but not equations of state. Our models do include semiconvection \citep[see][]{Kippy} but do not include overshoot or the diffusion of helium or heavy elements. We assumed a Big Bang helium value of Y=0.2485 at Z=0 \citep{Cyburt2004}, and did a linear fit to our solar value of Y=0.272683 at Z=0.018011 (so $Y=1.3426 Z+0.2485$). We convert Z to metallicity assuming $\log_{10}(\frac{Z/X}{Z_\sun/X_\sun})=[{\rm Fe/H}]$ for a solar mixture, with a correction factor for alpha enhanced models. We compare to other theoretical models in Section 3. In the initial part of this analysis, we use grids with a solar mixing length \citep{Bohm-Vitense1958} of 1.72, chosen to reproduce the solar luminosity ($3.827 \times 10^{33}$ erg s$^{-1}$) and the solar radius ($6.957 \times 10^{10}$ cm) at the solar age (4.57 Gyrs) \citep{Mamajek2012}.
For later parts of this study, we interpolate in a grid of models with mixing lengths of 1.22, 1.72, and 2.22. For the analysis in Section 3.2, we varied the initial helium abundance and surface boundary conditions in our model grid to explore their impact on the theoretical uncertainties. We ran models with fixed helium fractions of 0.239, 0.290, and 0.330 in addition to the solar calibrated value of 0.272683. We also created models with different atmospheric surface boundary conditions (defined as the boundary pressure at $\tau=2/3$): a look-up table from \citet{Kurucz1997} models and one from \citet{CastelliKurucz2004} models in addition to the gray atmosphere used in our main model grid.

We also compare to PARSEC models \citep{Bressan2012}. These models were run at metallicities from $-2.2$ to $+0.5$ and masses between 0.1 M$_\sun$ and 20 M$_\sun$ over that full metallicity range, with a mixture that is based on \citet{GrevesseSauval1998} with some updates from \citet{Caffau2011}. Our PARSEC grid does not include $\alpha$-element enhanced tracks. The PARSEC models use Irwin's FREEEOS\footnote{freeeos.sourceforge.net} tool to generate equations of state, OPAL high temperature opacities \citep{IglesiasRogers1996}, {\AE}SOPUS low temperature opacities \citep{MarigoAringer2009} and gray atmospheres. These models were run at a solar mixing length of $\alpha_{\rm MLT}=1.74$ \citep{Bohm-Vitense1958}. They include envelope overshoot as well as core overshoot in massive stars; microscopic diffusion is included in stars with a substantial surface convection zone unless they have a persistent convective core. Solar helium abundance is taken to be 0.276 and these models use $Y=0.2485+1.78 Z$. 

\section{Analysis} 
\subsection{Metallicity-Dependent Temperature Offset}
\begin{figure}
\begin{center}
\includegraphics[width=9cm, clip=true, trim=0.0in 0.8in 0.0in 0.8in]{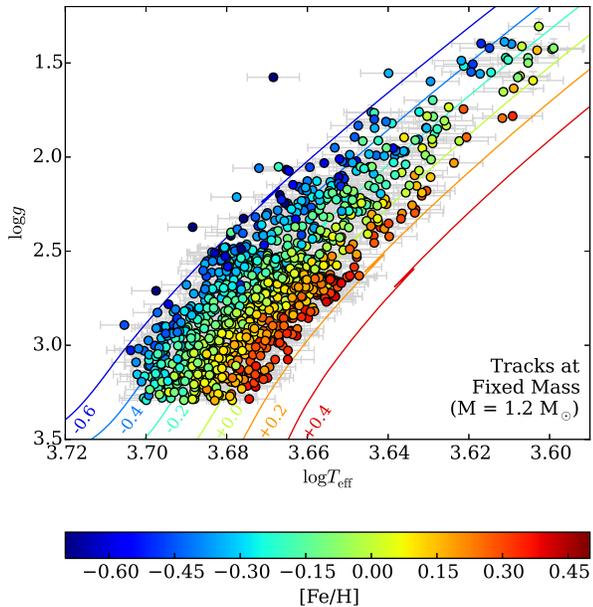}
\caption{The temperature and gravities of the stars in our sample between 1.1 and 1.3 M$_\sun$ are compared to models of 1.2 M$_\sun$ stars, close to the mean of this subsample. These 
tracks come from the YREC grid used in this work, and color coding indicates metallicity. The temperature offset is defined as the difference between the corrected APOGEE temperature and the temperature of a model with the measured mass, metallicity, [$\alpha$/Fe], and surface gravity, and is represented by the horizontal offset between a point and the corresponding line on this plot.}
\label{Fig:HRdiag}
\end{center}
\end{figure}

\begin{figure}
\begin{center}
\subfigure{\includegraphics[width=9cm]{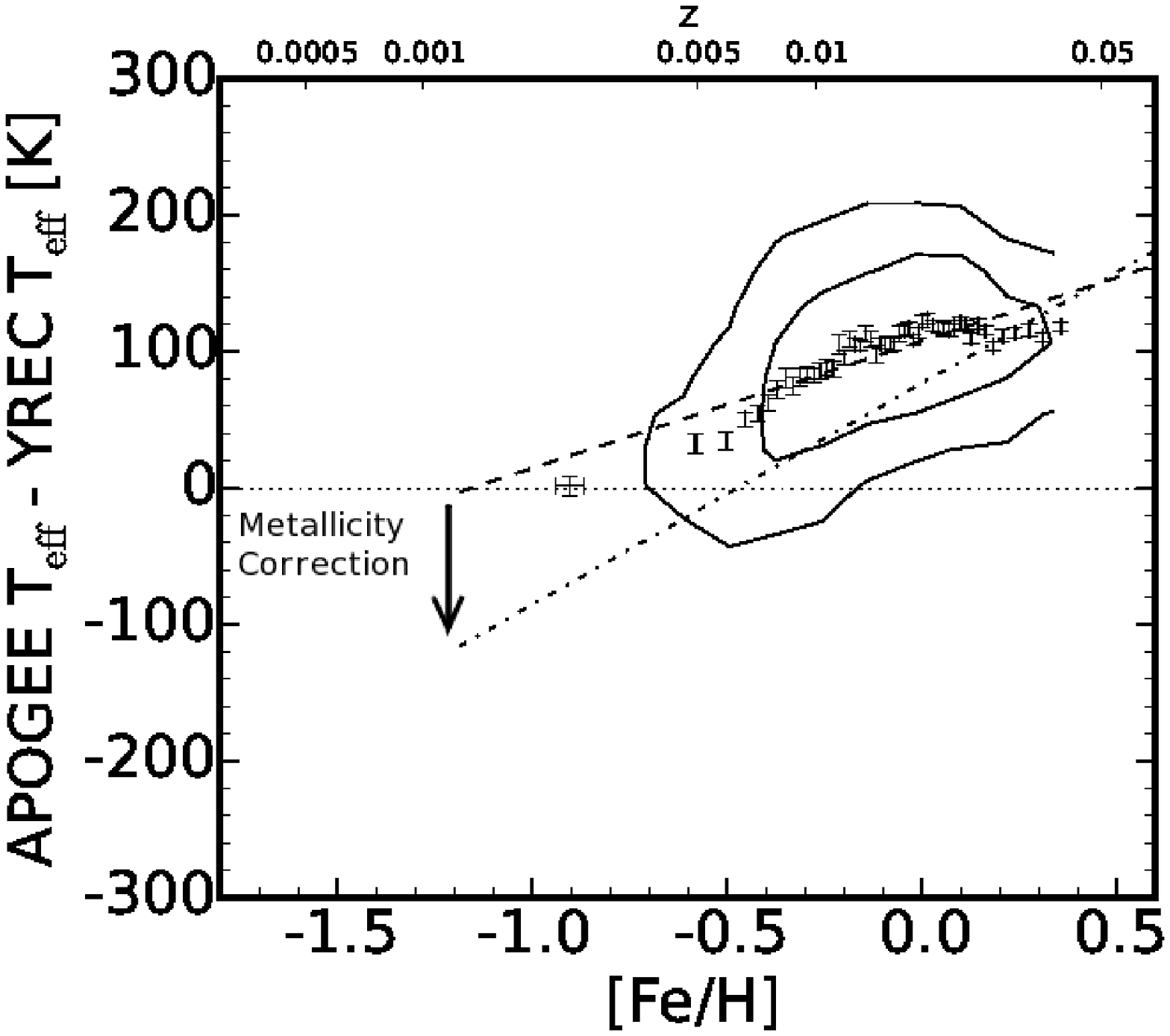}}
\subfigure{\includegraphics[width=9cm]{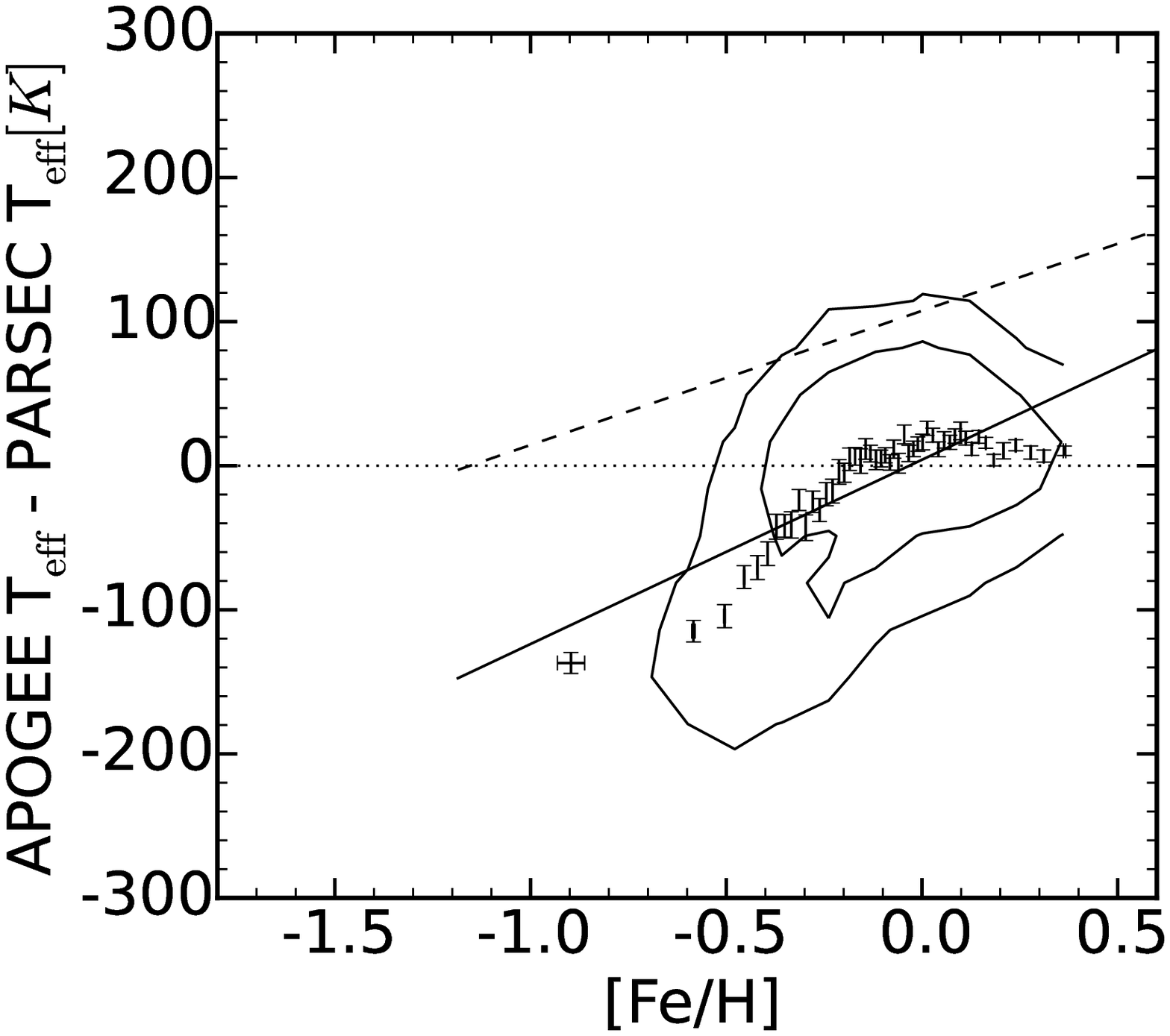}}
\caption{Plots of the difference between the temperature measured by APOGEE and the YREC or PARSEC model predictions as a function of metallicity. Contours indicate the  extent of 68 and 95 percent of the stars. The Z values on the top plot assume a solar mixture. Error bars indicate the mean and spread of binned data. The dot dashed line in the top panel indicates the slope we would have fit using the metallicity correction discussed in Section 2. We note that the temperature offset is best correlated with metallicity, with $\Delta T_{\rm eff, YREC}$ = 93.12 [Fe/H] + 107.50 K, shown as a dashed line in this and following figures.  The PARSEC models have a different normalization, but show a similar slope ($\Delta T_{\rm eff, PARSEC}$ = 127.88 [Fe/H] + 4.12 K). It is clear that a single linear fit does not completely capture the variation, and we discuss in the text regions consistent with a flat line. Note that we have terminated the lines at a metallicity of $-1.19$, as there are fewer than 10 stars in our sample below that metallicity.}
\label{Fig:temperature}
\end{center}
\end{figure}

 After interpolating to the appropriate helium for each metallicity, we perform a cubic spline interpolation in our model grid of mass, [Fe/H], [$\alpha$/Fe] and log g to infer a predicted $T_{\rm eff}$ for each star in our sample. This predicted $T_{\rm eff}$ is then subtracted from the APOKASC $T_{\rm eff}$ to compute a temperature offset. In Figure \ref{Fig:HRdiag} we show a subset of our data, chosen to have similar masses and differing compositions. The overlaid tracks represent theoretical expectations, and it is visually apparent that the temperatures seen in the data depend less on metallicity than predicted. To quantify this, we rank ordered the data in metallicity and averaged the differences between the predicted and actual temperatures in 64 star bins. In Figure \ref{Fig:temperature} we present these binned means as a function of metallicity, along with contours enclosing 68 and 95 percent of the sample. The differences show clear trends with metallicity that are highly statistically significant. These trends are present both for YREC (top) and PARSEC (bottom).
A linear fit gives $\Delta T_{\rm eff, YREC}= 93.1 {\rm [Fe/H]}+107.5$ K and $\Delta T_{\rm eff, PARSEC}= 127.9 {\rm [Fe/H]}+ 4.1$ K, with theoretical temperatures hotter than APOGEE observations at low metallicity and colder at high metallicity. We suspect that the steeper metallicity dependence of the PARSEC fit results from the use of a solar mixture without $\alpha$-element enhanced options. Correcting the solar mixture models for the effect of $\alpha$-element enhancement changes the temperatures by about 20 K on average \citep[see][]{Salaris1993}, and $\alpha$ enhancement correlates with metallicity. We also note that there are statistically significant deviations from a linear fit. For example, there is minimal evidence for a slope using only stars above a metallicity of -0.2. However, given the dependence of the slope on the details of the chosen metallicity scale, we caution against over-interpretation and provide only the linear fit. 
We note that the scatter around the linear relation for the YREC models is approximately Gaussian, with a standard deviation of 41 K, significantly less than the quoted APOGEE temperature uncertainties of 69 K \citep{Alam2015}. 

\begin{figure}
\begin{center}
\subfigure{\includegraphics[width=4cm]{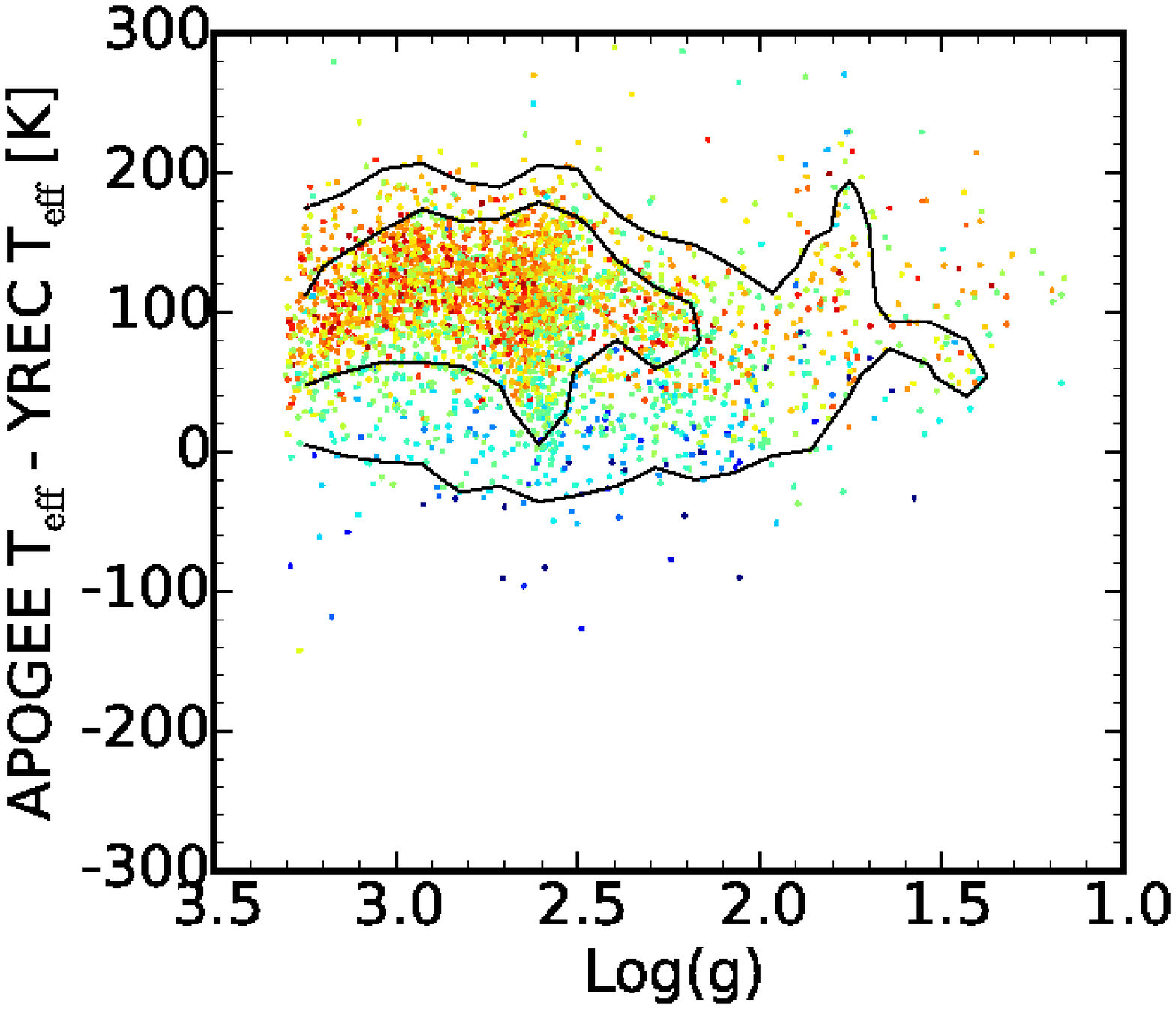}}
\subfigure{\includegraphics[width=4cm]{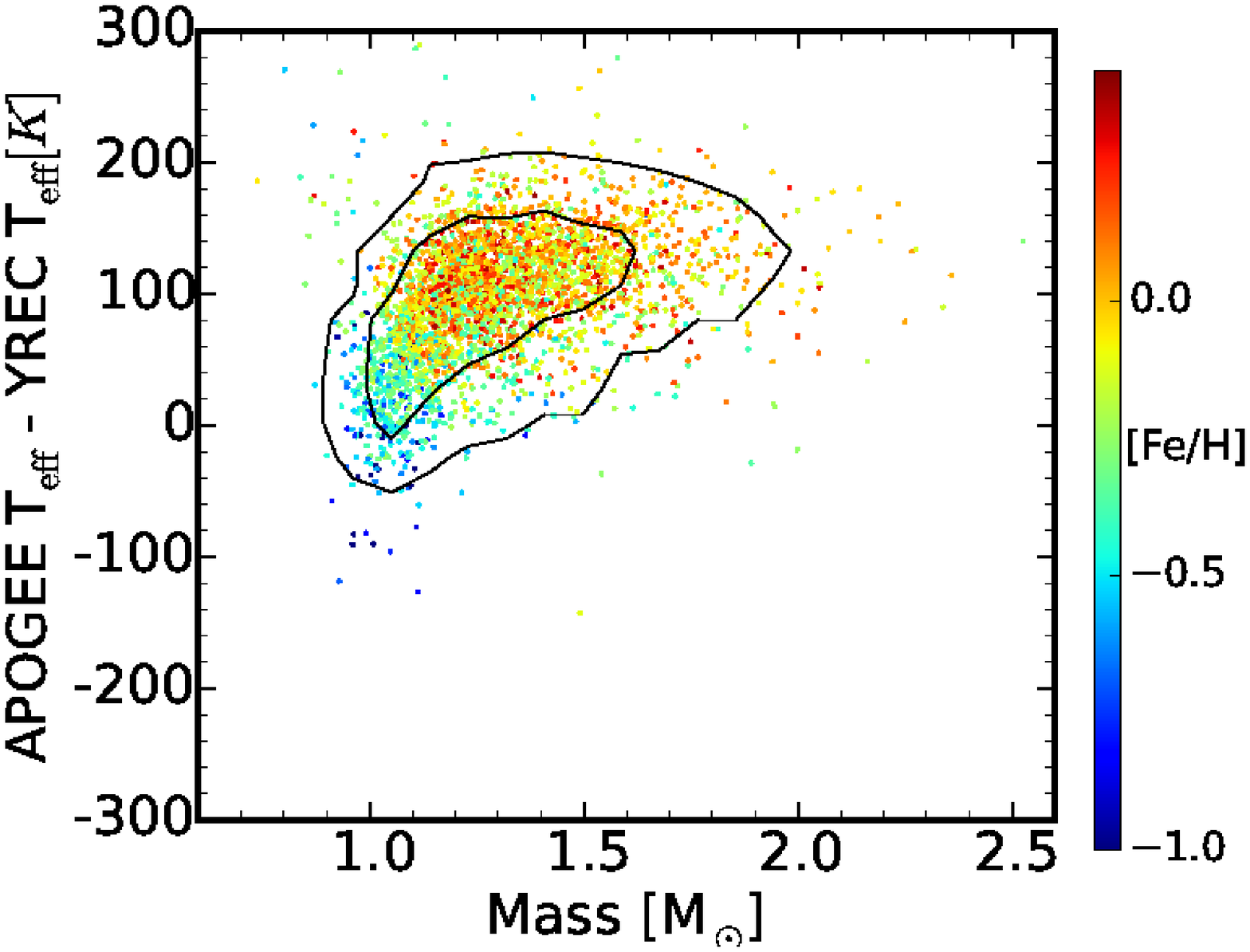}}
\subfigure{\includegraphics[width=4cm]{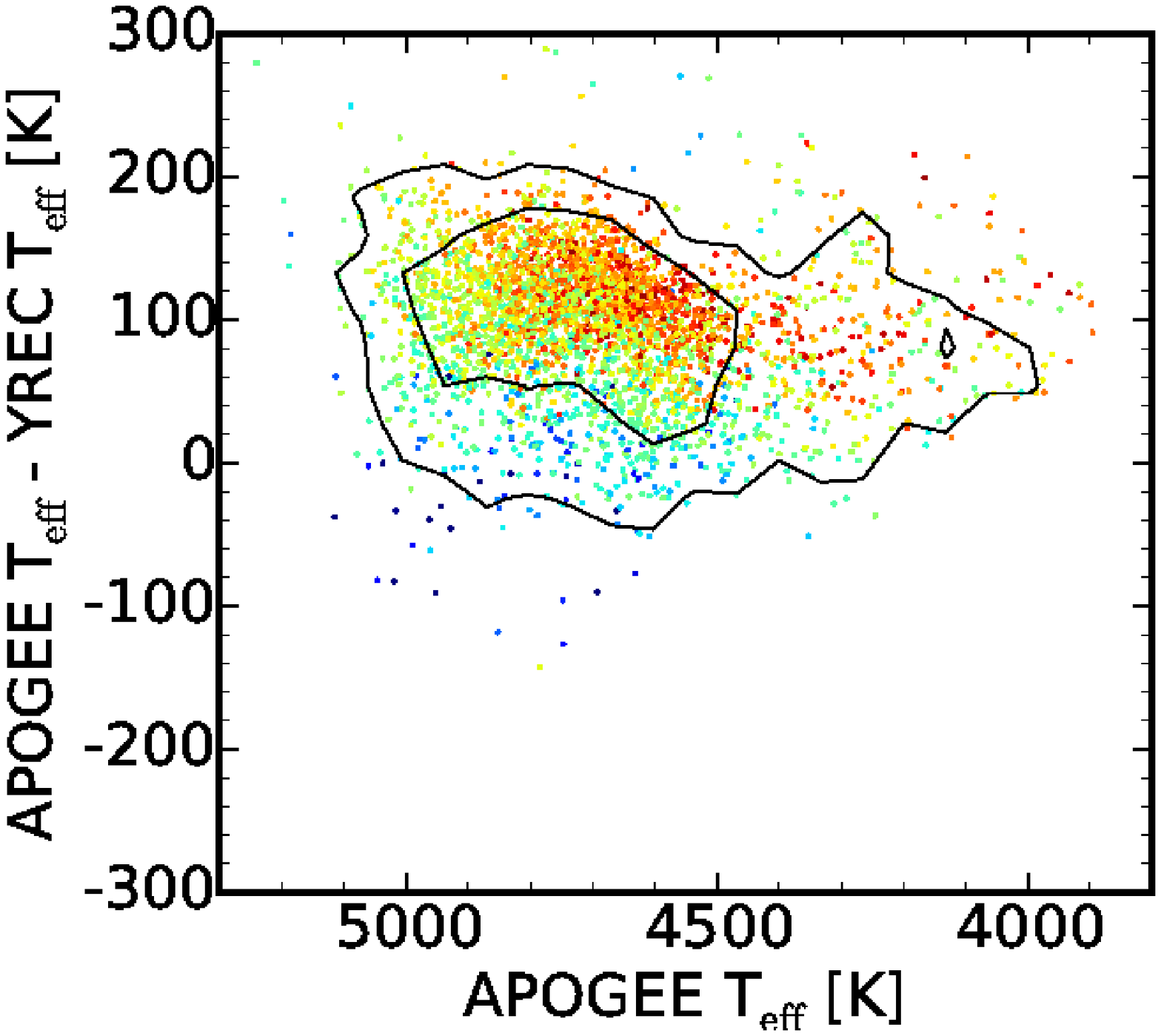}}
\subfigure{\includegraphics[width=4cm]{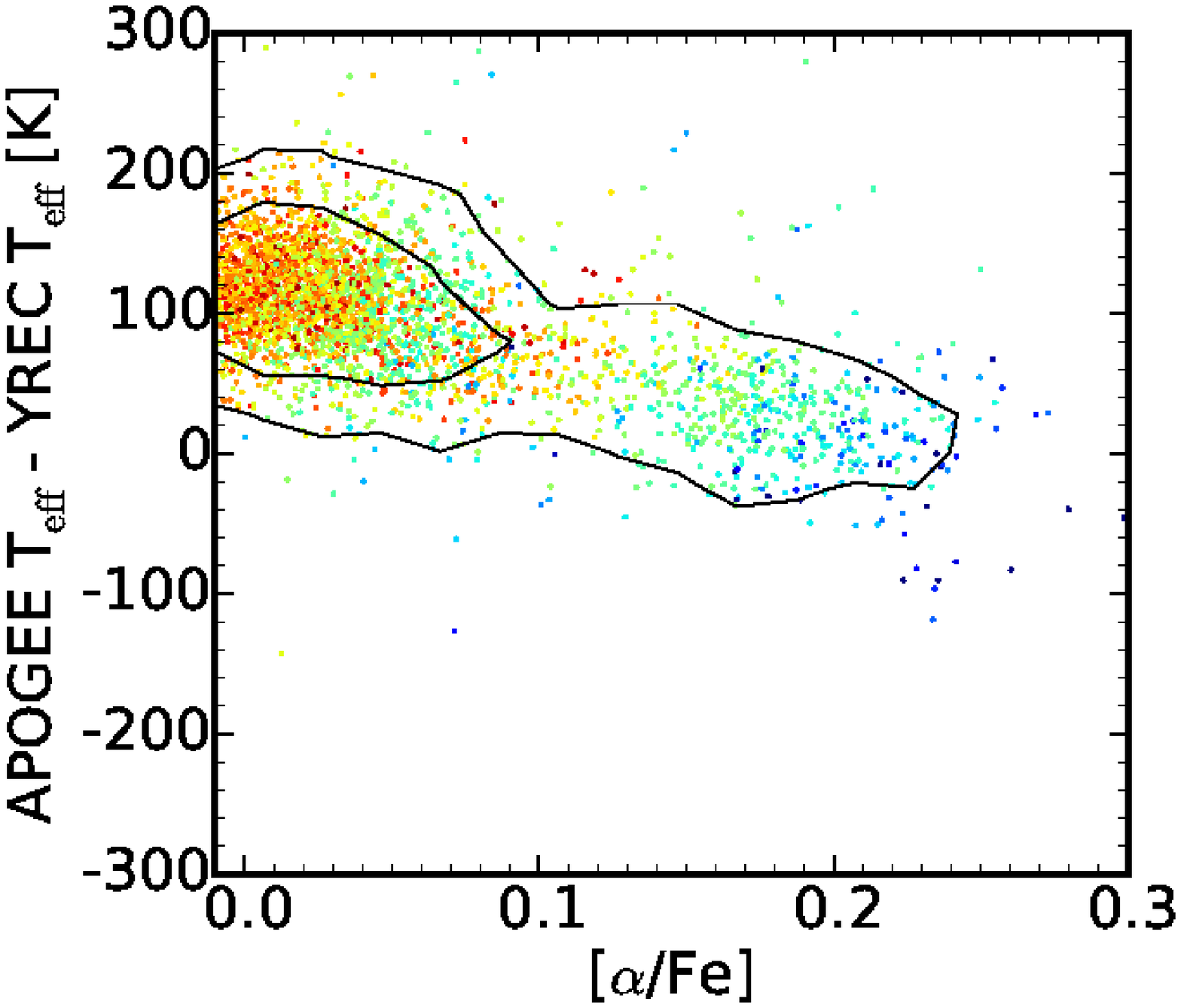}}
\caption{Plots of the difference between the temperature measured by APOGEE and the YREC model predictions as a function of various parameters, with log(g), mass, [$\alpha$/Fe] and  APOGEE temperature clockwise from top left. Points are color coded by metallicity, and contours indicate  the extent of 68 and 95\% of the sample. We note that the temperature offset is best correlated with metallicity (see Table \ref{Table:corrs}), although correlations appear on some of these plots as a result of  stellar evolution and galactic chemical evolution.}
\label{Fig:correlations}
\end{center}
\end{figure}

We also show in Figure \ref{Fig:correlations} the correlations between the temperature offsets and other parameters such as log(g), mass, temperature, and [$\alpha$/Fe]. We find that the temperature offset is best correlated with metallicity  (see Table \ref{Table:corrs} for the linear Pearson correlation coeffcients). We also see many of the correlations between the stellar properties that we would expect from stellar evolution (e.g. between temperature and gravity) and galactic chemical evolution (e.g. alpha enhanced stars are likely to be old, and therefore low mass and low metallicity).
The presence of the correlation between the temperature offset and the metallicity in two independent sets of models suggests that it may be a generic problem with evolutionary models. However, before we conclude that we must explore more mundane explanations. 

\begin{table}[htbp]

\caption{Linear Pearson correlation coefficients between the stellar properties (Mass, gravity, metallicity, $\alpha$ element enhancement, and temperature) and the temperature offset between the data and the YREC models ($\Delta T_{\rm eff}$). A value of 1 indicates a perfect positive correlation, a value of -1 would indicate a perfect negative correlation, and a value of 0 indicates no correlation. We show here that the temperature offset is best correlated with metallicity. }
\begin{tabular}{lrrrrrr}
\hline\hline
 & \multicolumn{1}{l}{Mass} & \multicolumn{1}{l}{log g} & \multicolumn{1}{l}{[Fe/H]} & \multicolumn{1}{l}{[$\alpha$/Fe]} & \multicolumn{1}{l}{$T_{\rm eff}$} & \multicolumn{1}{l}{$\Delta T_{\rm eff}$} \\ \hline
Mass & 1.00 & -0.03 & 0.31 & -0.51 & 0.12 & 0.12 \\ 
log g & -0.03 & 1.00 & 0.21 & -0.27 & 0.84 & 0.03 \\ 
\multicolumn{1}{l}{[Fe/H]} & 0.31 & 0.21 & 1.00 & -0.63 & -0.19 & 0.20 \\ 
\multicolumn{1}{l}{[$\alpha$/Fe]} & -0.51 & -0.27 & -0.63 & 1.00 & -0.21 & -0.15 \\ 
$T_{\rm eff}$ & 0.12 & 0.84 & -0.19 & -0.21 & 1.00 & 0.00 \\ 
$\Delta T_{\rm eff}$ & 0.12 & 0.03 & 0.20 & -0.15 & 0.00 & 1.00 \\ 
\end{tabular}
\label{Table:corrs}
\end{table}

\subsection{ Model Uncertainties}

\begin{figure}
\begin{center}
\subfigure{\includegraphics[width=8cm]{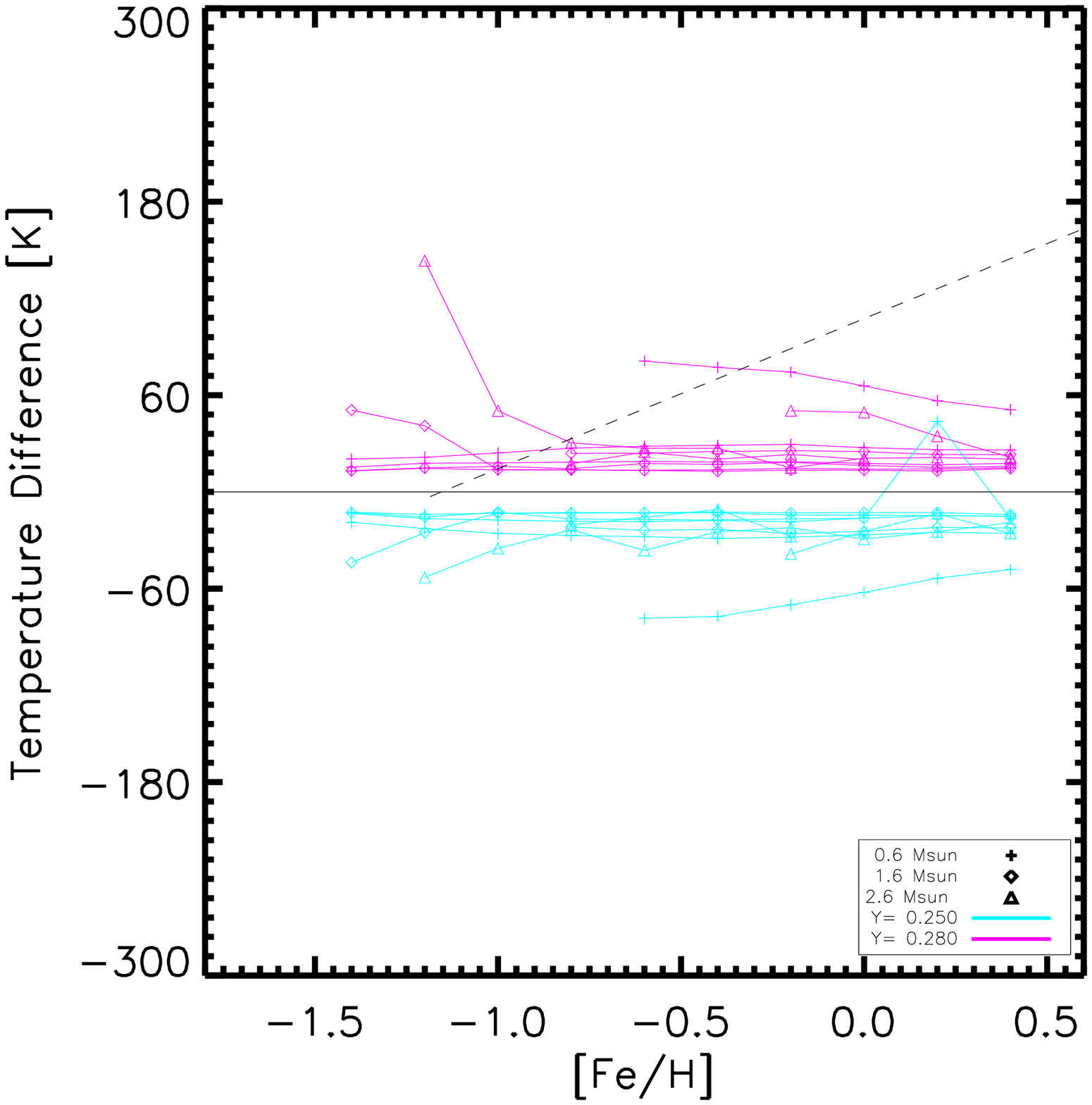}}
\subfigure{\includegraphics[width=8cm]{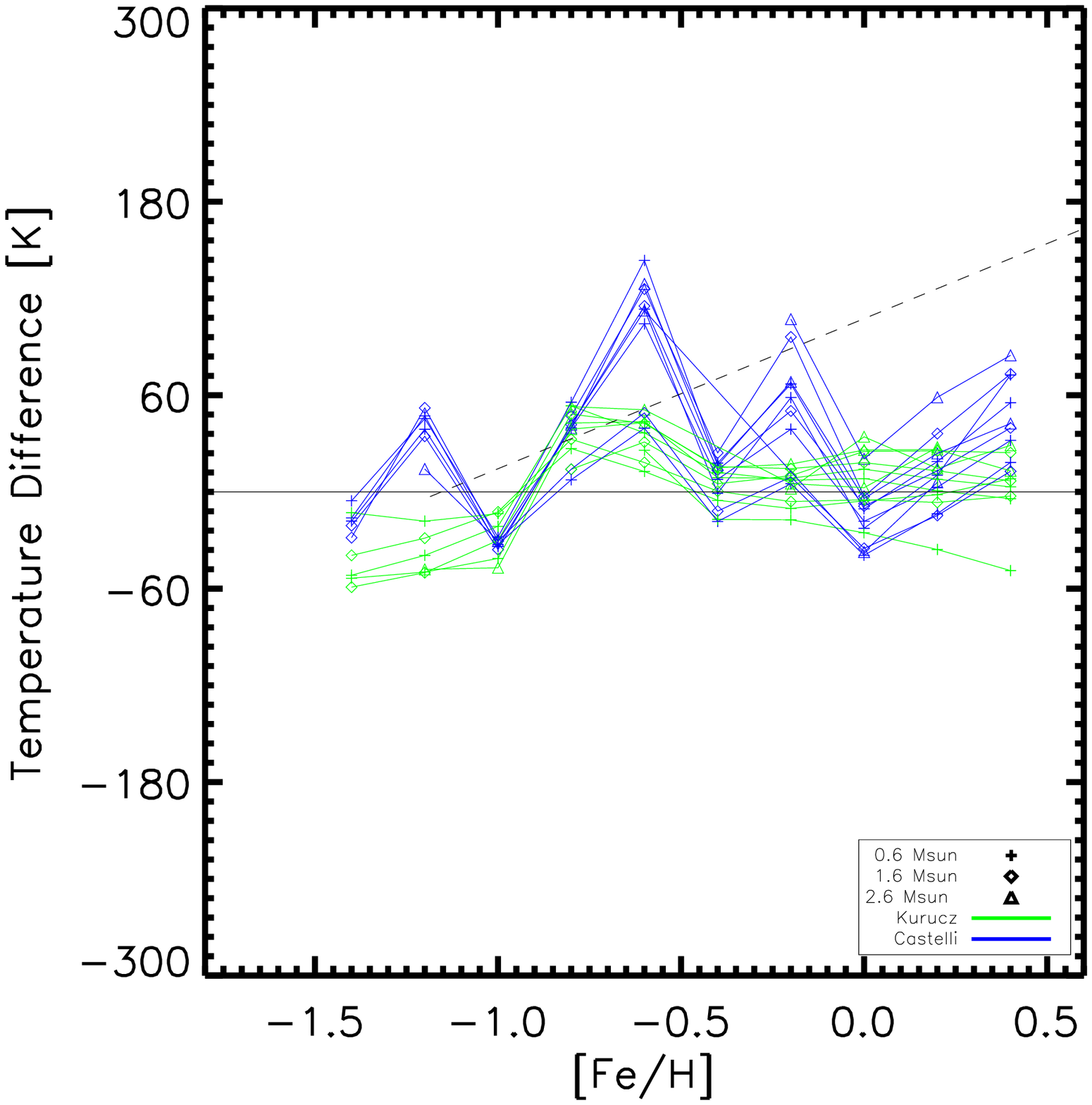}}
\caption{Plots of the difference in model temperature over a range of masses and gravities as a function of metallicity for different helium values (top) and atmosphere boundary conditions (bottom). The mean trend line of temperature offset versus metallicity from Figure  \ref{Fig:temperature} is again shown here as a dashed line. 
}
\label{Fig:theory}
\end{center}
\end{figure}

First, we explore whether known uncertainties in the stellar model physics, such as helium content and atmosphere boundary condition, can cause shifts in giant branch temperature on the level of what is observed in the APOKASC data. We show in Figure \ref{Fig:theory} the expected scale shifts, as well as the differential offsets with metallicity, caused by choosing a different atmosphere boundary condition (Kurucz or Castelli), or by assuming a different helium abundance.
The mean trend in the APOKASC data is shown as a dashed line. It is clear that while there are many uncertainties in the theoretical models that can cause changes in the temperature of the giant branch locus, these shifts rarely show a strong metallicity dependence and are less than 30 K dex$^{-1}$, not large enough to explain our offset, which is about three times larger.

\subsection{Mass Biases}
\begin{figure}
\begin{center}
\subfigure{\includegraphics[width=4cm]{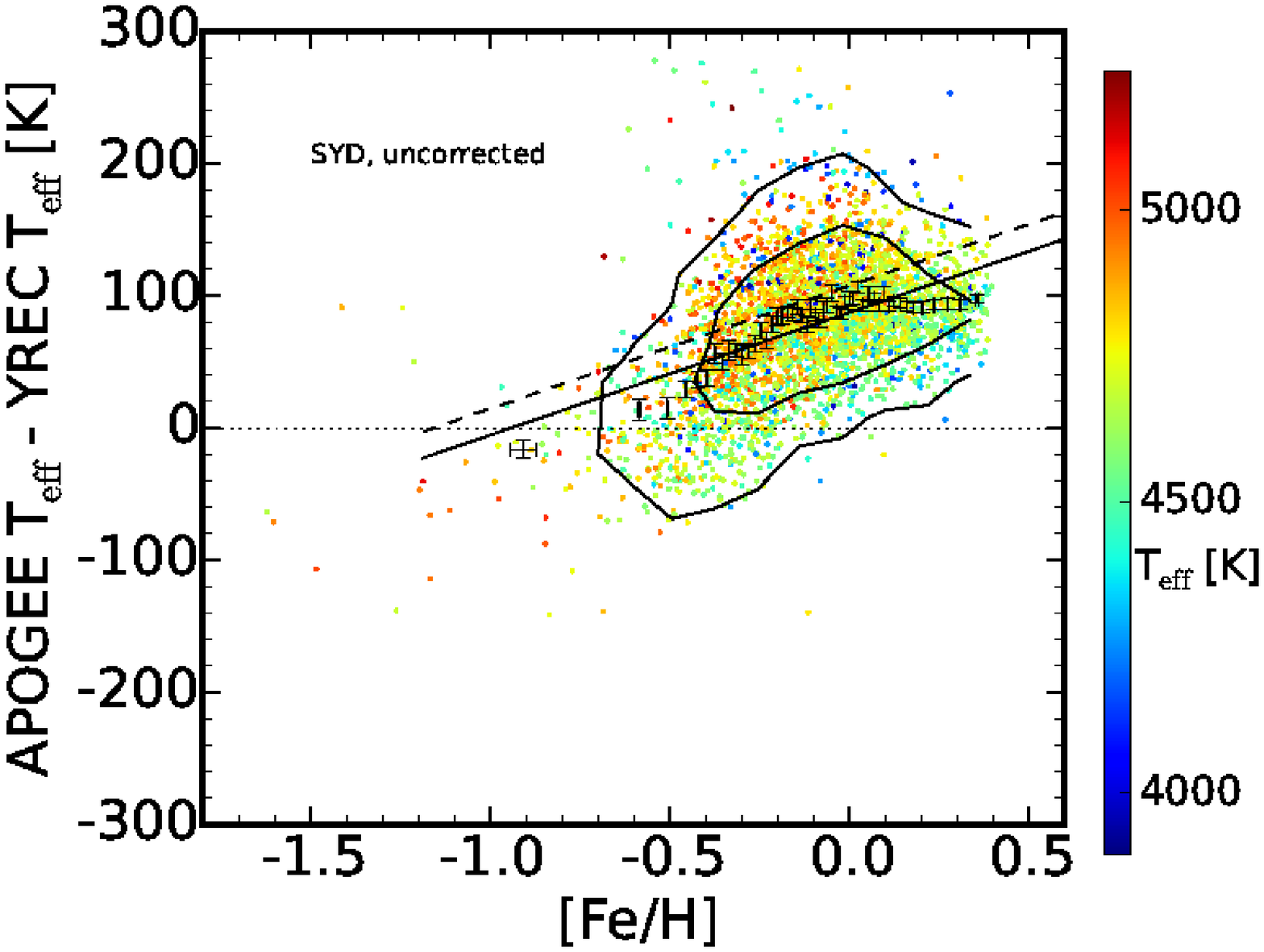}}
\subfigure{\includegraphics[width=4cm]{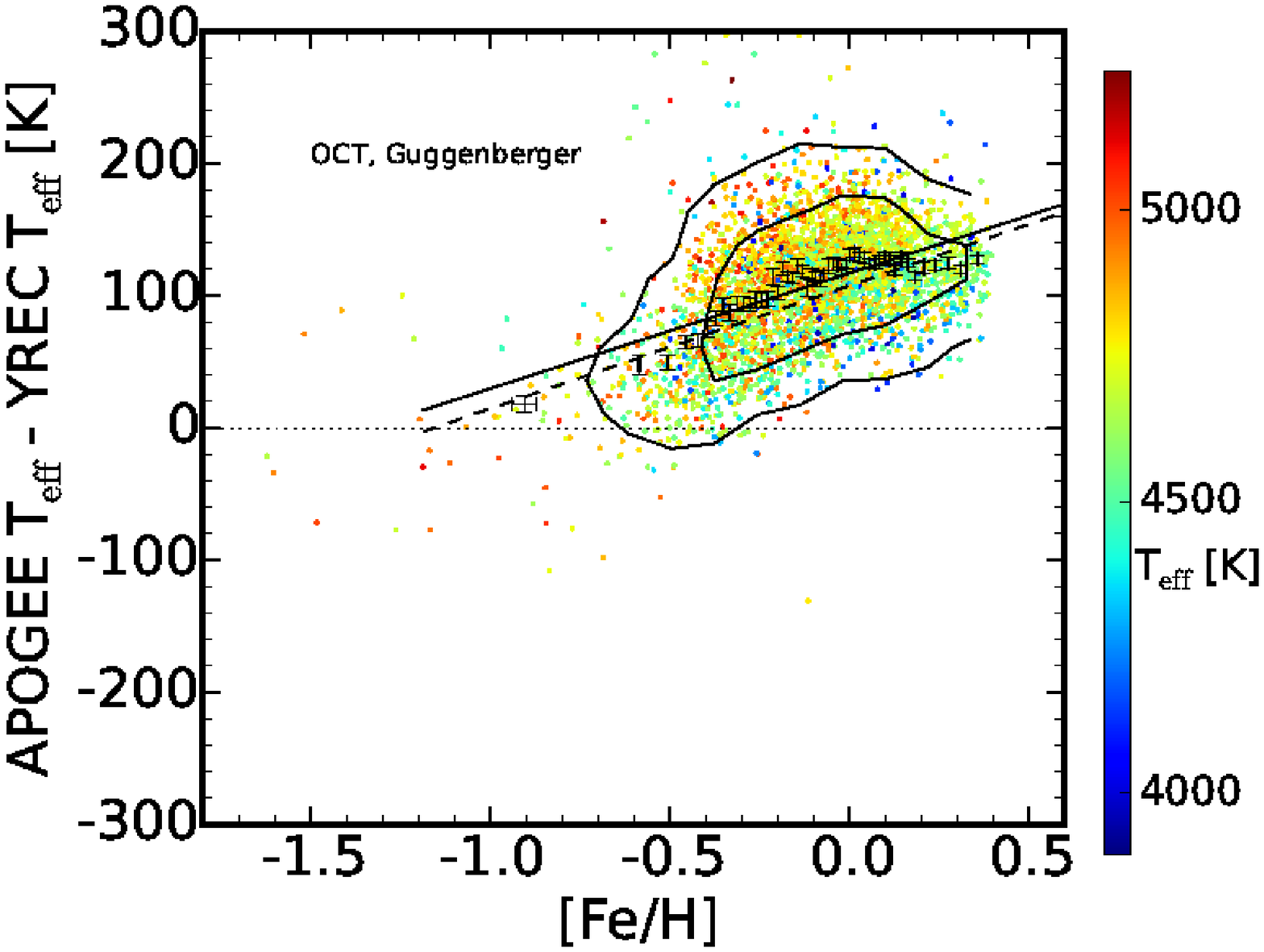}}
\subfigure{\includegraphics[width=4cm]{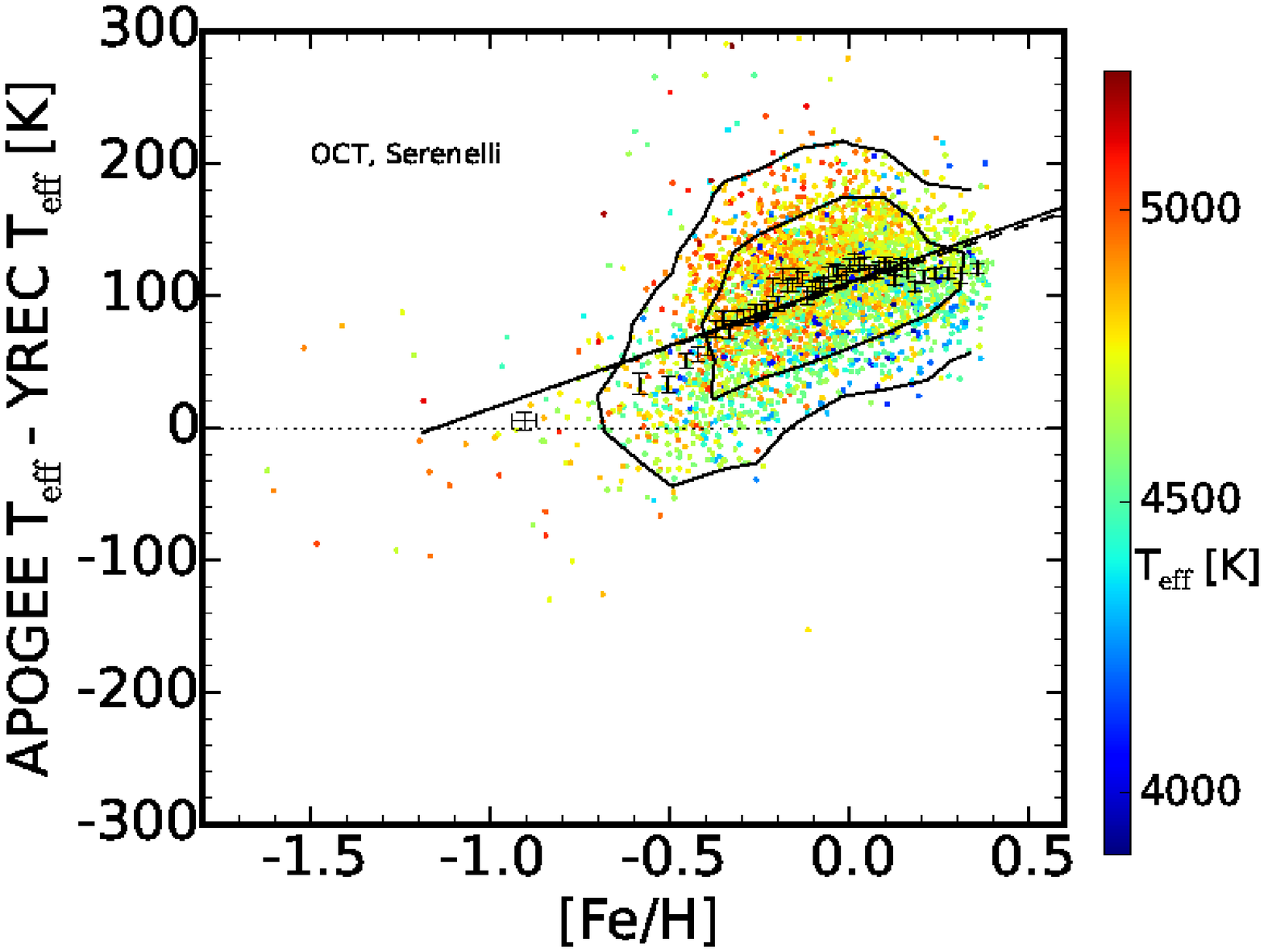}}
\subfigure{\includegraphics[width=4cm]{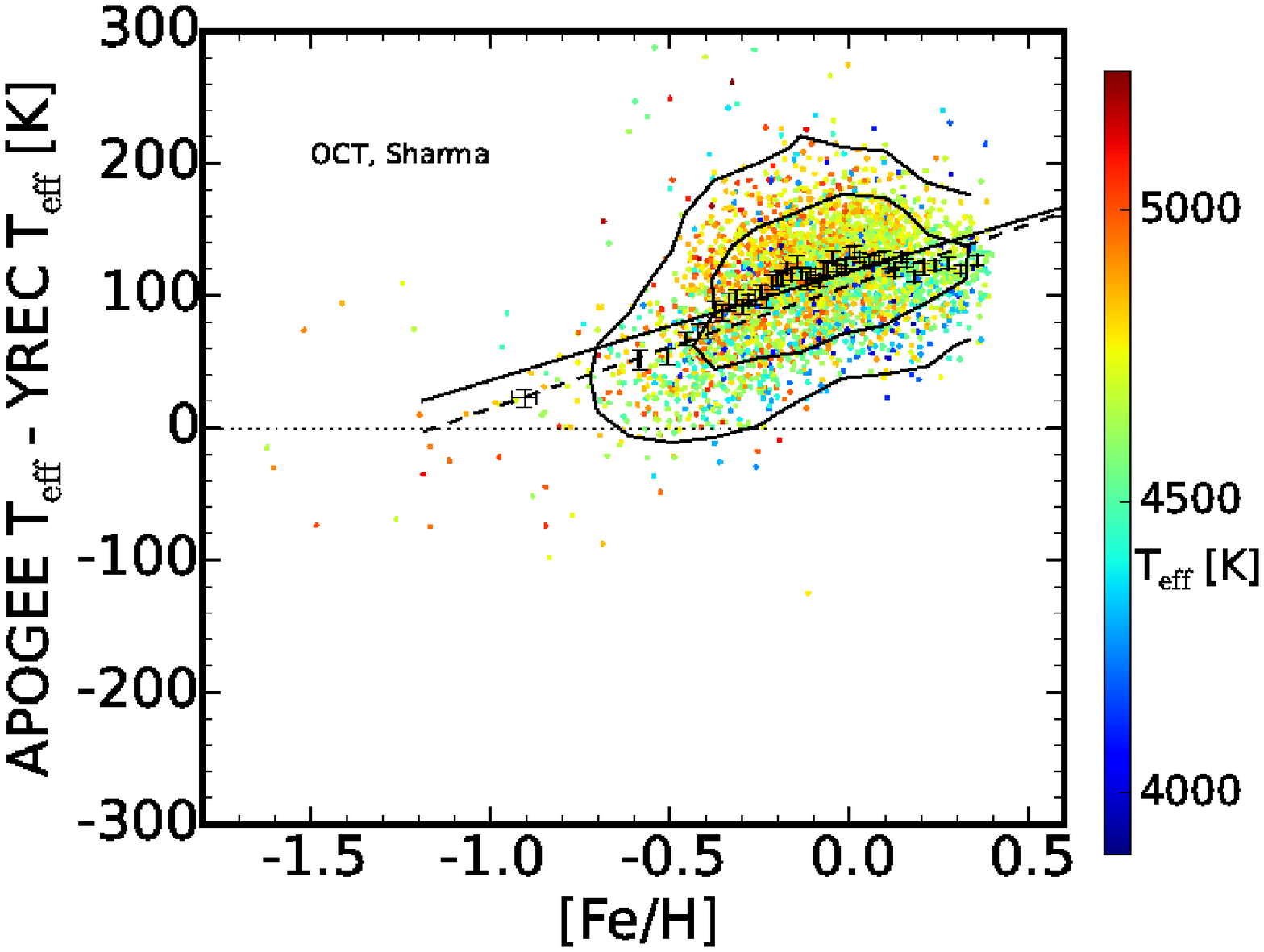}}
\caption{Plots of the offset between the APOGEE temperature and the model temperature for four different mass scales. The top left plot uses scaling relation masses and gravities from the SYD pipeline \citep{Huber2009} rather than the OCT pipeline. The top right panel uses masses computed from the OCT pipeline values, but with corrections as given in \citet{Guggenberger2016}. The bottom left panel uses masses computed using grid based modeling as in Serenelli et al. (2016, in prep) using the actual APOGEE temperature errors. The bottom right panel uses masses computed using the \citet{Sharma2016} corrections. The temperature offset persists no matter the choice of seismic mass scale. The dashed lines indicate the best fit from Figure \ref{Fig:temperature}; solid lines indicate the best fit for each panel. Contours indicate the extent of 68 and 95 percent of the sample.}
\label{Fig:masses}
\end{center}
\end{figure}

Second, we explore whether this trend could be due to systematic errors in the APOKASC data. While the seismic masses we use are extremely precise, there have been suggestions that they need to be corrected directly for the effects of metallicity and temperature \citep{White2011, Epstein2014,Sharma2016, Guggenberger2016}. We tested whether the observed correlation between temperature offset and metallicity is a result of using uncorrected scaling relations to determine the mass. For this we estimated masses from (1) grid-based modeling using 
$\Delta \nu$ computed from radial mode eigenfrequencies and the scaling relation for $\nu_{\rm max}$ (Serenelli et al. 2016, in prep), (2) masses computed using the scaling relations but with the grid-interpolated correction to the $\Delta \nu$ relation based on radial modes by \citet{Sharma2016}, and (3) scaling relation masses using the \citet{Guggenberger2016} analytical correction to the $\Delta \nu$ relation. We find that corrections generally have an effect less than 0.1 M$_\sun$ (less than 30 K), and that these corrections do not substantially affect the metallicity dependence of the temperature offsets (see Figure \ref{Fig:masses}). We also show that using a different procedure to measure the global seismic parameters \citep[SYD, ][]{Huber2009} also does not substantially affect the results.

We note that asteroseismic mass estimates are sometimes adjusted to better agree with the properties expected from a grid of stellar models; this is referred to as grid-based modeling. However, we caution that grid-based modeling does not necessarily improve asteroseismic results if the underlying tracks are not correctly located. In the red giant case, naive grid-based modeling will change the inferred posterior temperatures to reduce the discrepancy between the observed properties and the model predictions, dragging the metal poor stars to hotter posterior temperatures and metal rich stars to cooler posterior temperatures. These changes average 50-100 K at [Fe/H]=$\pm$0.5. If the quoted temperature uncertainties are reduced to prevent this from occurring, the grid based modeling will instead alter the masses by about $\pm$ 0.2 M$_\sun$ at [Fe/H]=$\pm$0.5 to reduce the discrepancy between the data and the models. These changes are completely consistent with our expectations given the metallicity dependent temperature offset we discuss above, and we therefore chose not to use grid modeling based masses in this work as it complicates the signal we are examining.  

\subsection{Spectroscopic Uncertainties}
In this analysis, we use the updated APOGEE temperatures, which have been corrected to the photometric scale. There should not, therefore, be correlated errors between the temperatures and the metallicities measured by the APOGEE ASPCAP pipeline. 
We did however test whether substantial changes would occur in the measured temperatures and metallicities if we used the APOGEE grid of MARCS atmospheres, rather than the Kurucz atmospheres used in DR13, for the 27 stars in our sample with $T_{\rm eff} < 4400$ K, where the MARCS grid is available \citep[see][]{Zamora2015}. We find that this on average moves the temperatures about 12 K cooler and makes a star about 0.007 dex more metal poor. However, these changes are small and not strongly metallicity dependent; choice of spectroscopic model atmosphere is thus unlikely to contribute to our observed offset.

We have also checked that the temperature offset is still present in the \citet{Hawkins2016} reanalysis of the previous APOKASC data set. This gives us more confidence that it is not some peculiarity of the ASPCAP analysis that is causing the metallicity dependent temperature offset.

One other major source of uncertainty in the spectroscopic parameters is the use of atmosphere models constructed assuming local thermodynamic equilibrium (LTE). It is well known that accounting for the effects of non-local thermodynamic equilibrium (NLTE) can substantially change the measured stellar parameters \citep[see e.g.][]{Asplund2005b}. Investigations by \citet{Ruchti2013}, for example, indicated that using a NLTE analysis of optical iron lines could change the measured temperature, metallicity, and gravity in a metallicity dependent way. While there have been fewer studies of the effects of NLTE corrections to lines in the infrared \citep{Zhang2016}, and it is unclear how the full spectral fitting done for the APOGEE spectra would differ from the analysis of individual lines, we use the \citet{Ruchti2013} study to estimate the kind of parameter changes we might expect if NLTE effects were included. In our analysis, we are calibrating our temperatures to the photometric scale, which should be unaffected by spectroscopic systematics like NLTE effects. Similarly, since we are using seismic gravities, we are not concerned about the effect of NLTE corrections on the measured surface gravity. In terms of metallicity, which has not be calibrated to a fundamental scale, the \citet{Ruchti2013} analysis indicates that the measured metallicities change by less than 0.1 dex for high metallicity stars like our APOKASC sample. We therefore choose to proceed with the assumption that our measured trend is not caused by spectroscopic uncertainties, but recommend further investigation into the effect of NLTE corrections on the APOGEE analysis. 

\subsection{Comparison with Other Datasets}
\subsubsection{Star Clusters}

\begin{figure}
\begin{center}
\subfigure{\includegraphics[width=9cm,  clip=true, trim=0cm 0cm 3cm 0cm]{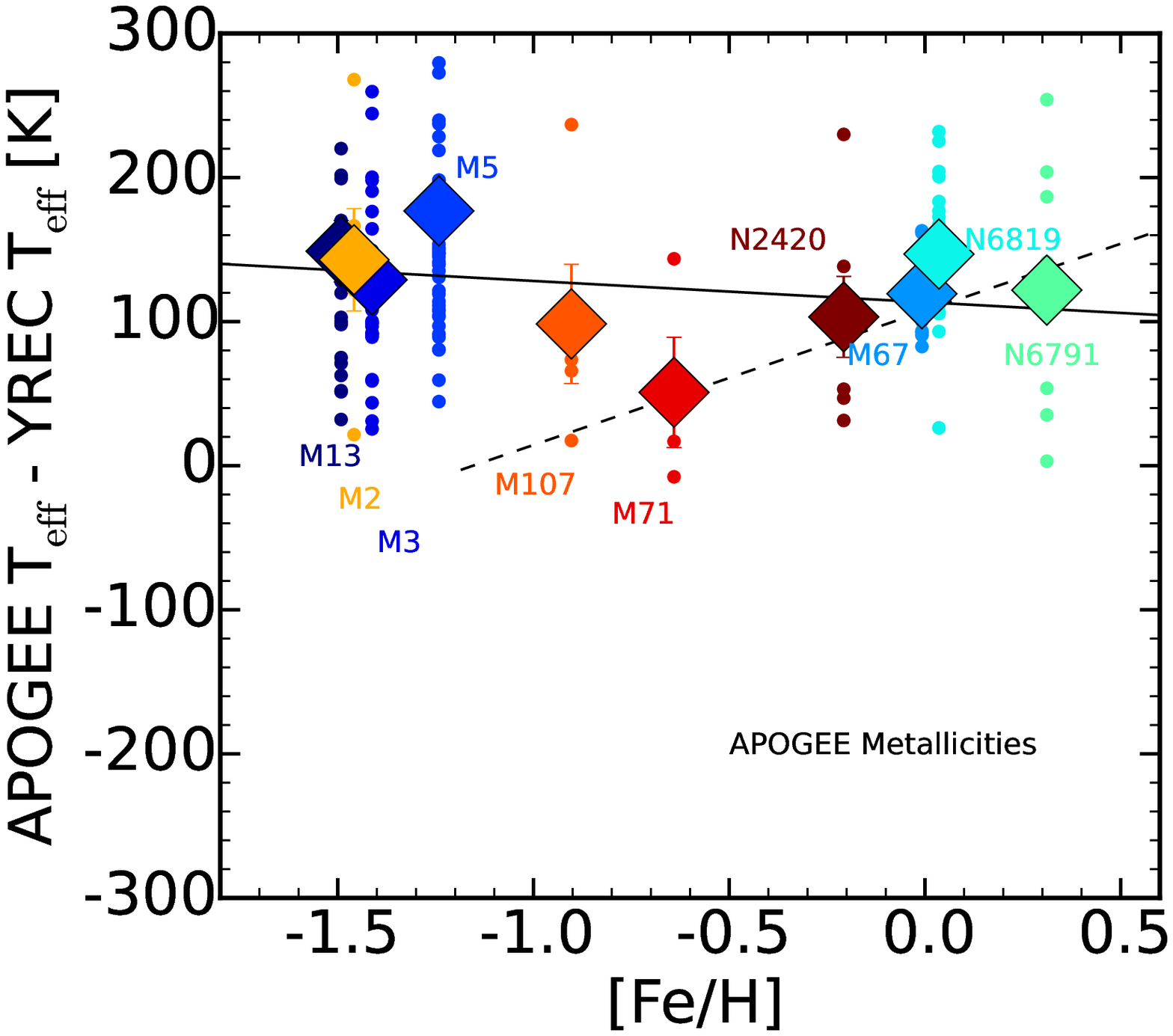}}
\subfigure{\includegraphics[width=9cm,  clip=true, trim=0cm 0cm 3cm 0cm]{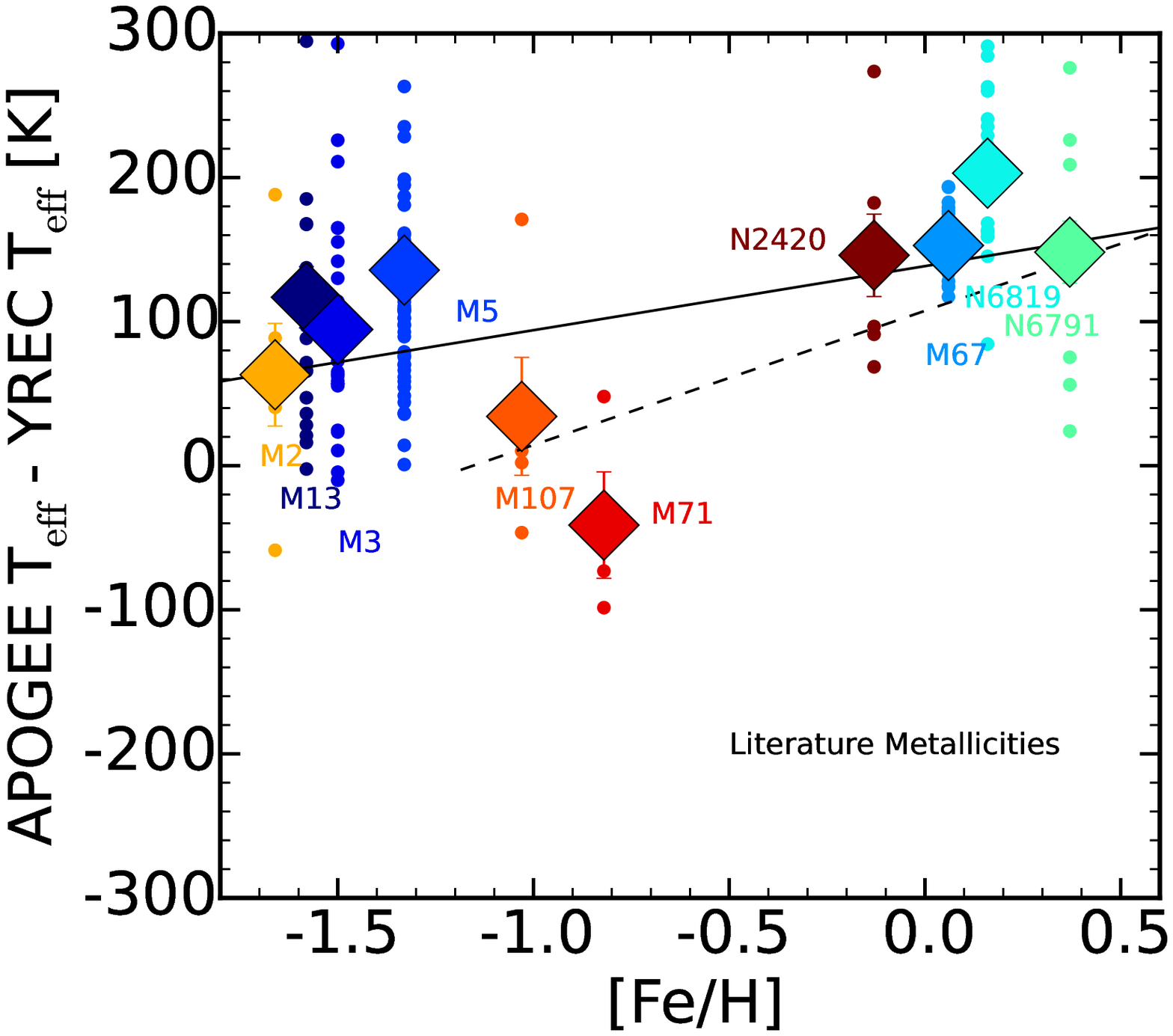}}
\caption{Plot of the offset between the measured APOGEE and model temperatures for stars in clusters, using APOGEE metallicities in the top plot and literature metallicities from \citet{Holtzman2015} in the bottom plot. Large diamonds indicate cluster means, and show the approximate size of the systematic error bars given a mass uncertainty of 0.05 M$_\sun$ and a metallicity uncertainty of 0.05 dex. Error bars indicate the standard error on the mean if it is larger than the systematic uncertainties. Individual stars are represented by smaller points. We note that many of these clusters were also used to check the APOGEE temperature calibration (red stars in Figure \ref{Fig:teffcor}), so they are consistent with the photometric temperature scale. The solid line indicates the best fit in each plot, treating each cluster as a single point. The dashed line represents the mean trend with metallicity from Figure \ref{Fig:temperature}, note that it is consistent with clusters in the metallicity range spanned by the APOKASC data, but inconsistent with the low metallicity globular clusters (see text). We therefore caution against extrapolating these results to the low metallicity regime. Because of the known uncertainties with the APOGEE metallicity scale (see Section 2) we show this figure with both the APOGEE (top) and literature metallicity scales (bottom).} 

\label{Fig:cluster}
\end{center}
\end{figure}

Star clusters are a natural laboratory for checking the concordance between theoretical models and stellar data.  In principle, since clusters are coeval and the red giant branch lifetime of a star is relatively short, the red giants in a cluster should represent a homogeneous sample of a single known mass and composition. 
The offset between the cluster isochrone and the measured temperature at the gravity of each star can therefore be used to measure a temperature offset similar to our procedure for asteroseismic data.  This approach is powerful but caution is needed.  Some methods for determining color-temperature relationships, for example, or fitting for cluster parameters, involve minimizing deviations between isochrone and data (for example, by applying zero point offsets to align cluster CMDs with theory).  Globular clusters are also not simple stellar populations, and deviations from a universal mix may not be captured in theoretical isochrones assuming a solar mixture with alpha enhancement \citep{Beom2016}.  With these caveats in mind, there is a large set of APOGEE data for star clusters, and we have used it to test both our temperature scale (Figure \ref{Fig:teffcor}) and the offsets between cluster data and theory (Figure \ref{Fig:cluster}).

For globular clusters, we took cluster members with APOGEE data from M2, M3, M5, M13, M71, and M107 as described in \citet[][2016, in prep.]{Holtzman2015}.  We note that this list excludes clusters below a metallicity of -2.0 because of known uncertainties with the spectroscopic gravity corrections in such clusters. We also included only stars with first generation abundance patterns, as the ASPCAP fitting process \citep{GarciaPerez2015} yields spurious correlated metallicity and temperature offsets for stars whose base model atmospheres differ significantly from the assumed base model.  For the open cluster NGC 2420 we took the members from \citet{Souto2016}; M67 members were taken from stars with asteroseismic and APOGEE data as tabulated in Stello et al. (2016, submitted).  Finally, for the open clusters NGC 6819 and 6791 we took targets with asteroseismic and spectroscopic data from Pinsonneault et al. (2016, in prep).  In open clusters where evolutionary states where available, we used the relevant analysis to exclude red clump stars. In the globular clusters, we used the red clump cut defined by \citet{Bovy2014} to limit our sample to shell burning giants. We excluded stars whose measured APOGEE metallicity was more than three sigma away from the cluster mean, as well as stars outside the gravity range of our APOKASC sample (log(g) between 3.3 and 1.1). This limits contamination by AGB stars, and also eliminates any concern that the temperature offset may become gravity dependent at very low gravities. 

Metallicities and $\alpha$-element enhancements reflect the mean abundances in the chosen samples (see Table \ref{tab:cluster}).  In the cases of M67, NGC 6819 and NGC 6791 red giant masses can be accurately predicted by eclipsing binary stars with measured masses on the upper main sequence.  We assume a red giant branch mass of 1.63 M$_\sun$ for  NGC 2420, consistent with \citet{Souto2016} 
A red giant branch mass of 0.85 M$_\sun$ was assumed for the old globular clusters.  We note that the temperature offsets on the RGB are relatively insensitive to a change in mass ($<$10 K per 0.05 M$_\sun$), but they are more sensitive to a change in metallicity ($\sim$25 K per 0.05 dex). Given our sample sizes, these systematic errors are in most cases more significant than random errors, and their quadrature sum is approximately represented by the size of the diamonds in Figure \ref{Fig:cluster}.  We assumed that all cluster members had the mean metallicity and [$\alpha$/Fe] of the cluster and computed the offset between an evolutionary track with the cluster RGB mass and metallicity and the individual stellar spectroscopic temperatures at the corrected spectroscopic gravity; the spectroscopic temperatures were corrected for the trends described in Section 2.  The results are illustrated in Figure \ref{Fig:cluster}.  Using the cluster data with APOGEE metallicities, we see no strong evidence for a trend across the full metallicity range, although there is some evidence that the intermediate metallicity globular clusters are less offset from the tracks than the lower metallicity clusters. However, if we use literature metallicities from \citet{Holtzman2015}, we see a trend that averages about 45 K per dex across the full metallicity range, with clusters in the APOKASC metallicity range (above [Fe/H] $\sim -$1) consistent with the trend seen in the APOKASC data. 
Given how dependent this analysis is on the fundamental metallicity scale, something this paper is not equipped to address, we tentatively conclude that the cluster sample is not inconsistent with our trend in the metal-rich domain, but that our data should not be extrapolated into the metal-poor domain without further work on the absolute metallicity calibration of APOGEE in that domain.

\begin{table}[htbp]
\begin{center}
\caption{Cluster properties used in this paper. Literature metallicities come from \cite{Holtzman2015}.}
\label{tab:cluster}
\begin{tabular}{lrrrr}

Cluster & \multicolumn{1}{l}{Mass} & \multicolumn{1}{l}{[Fe/H]} & \multicolumn{1}{l}{[$\alpha$/Fe]} & \multicolumn{1}{l}{Lit. [Fe/H]} \\ \hline
M13 & 0.85 & -1.49 & 0.22 & -1.58 \\ 
M2 & 0.85 & -1.46 & 0.23 & -1.66 \\ 
M3 & 0.85 & -1.41 & 0.18 & -1.50 \\ 
M5 & 0.85 & -1.24 & 0.21 & -1.33 \\ 
M107 & 0.85 & -0.90 & 0.29 & -1.03 \\ 
M71 & 0.85 & -0.64 & 0.18 & -0.82 \\ 
N2420 & 1.63 & -0.21 & 0.01 & -0.13 \\ 
M67 & 1.36 & -0.01 & 0.00 & 0.06 \\ 
N6819 & 1.63 & 0.04 & 0.00 & 0.16 \\ 
N6791 & 1.15 & 0.31 & 0.09 & 0.37 \\ 
\end{tabular}
\end{center}

\end{table}

\subsubsection{Hipparcos Giants}
\begin{figure}
\begin{center}
\includegraphics[width=9cm]{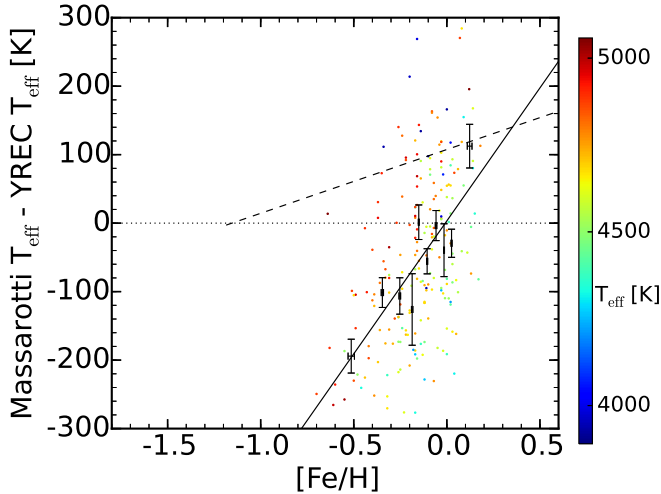}
\caption{Plot of the offset between the measured and model temperatures for the \citet{Massarotti2008} spectroscopic sample of Hipparcos giants. Error bars represent binned data. The solid line indicates the best fit to the sample, with $\Delta T_{\rm eff}$ = 388.3 [Fe/H] + 3.29 K. The dashed line is our fit to the APOKASC data. We suspect that the difference in slope could be due to the heterogeneous derivation of properties in the \citet{Massarotti2008} work;  the offset in the mean temperature difference is likely due to calibration choices.}
\label{Fig:mas2008}
\end{center}
\end{figure}

Another data set with all the parameters necessary for our analysis is the \citet{Massarotti2008} sample of red giants. These stars have distances from {\it Hipparcos} data as well as luminosities computed from their V-band absolute magnitudes combined with bolometric corrections from \citet{VandenBergClem2003}. They do have heterogeneously determined temperatures (photometric), metallicities (mostly spectroscopic), and gravities (partially spectroscopic, partially determined through comparison with isochrones). The combination of these parameters allows the calculation of the mass of each star independently of stellar models, and these stars can then be analyzed in exactly the same manner as our APOKASC stars. Their temperature offsets versus metallicity are shown in Figure \ref{Fig:mas2008}. We use only stars in the same mass range as the APOKASC sample, 0.6 to 2.6 M$_\sun$, and use the temperature and metallicity dependent gravity cuts from \citet{Bovy2014} to exclude red clump stars in the Massarotti sample, leaving 173 stars. Because these stars only have measured bulk metallicities rather than $\alpha$-element enhancements, we assume that all stars in the sample have [$\alpha$/Fe] = 0. Knowing that low-metallicity stars are more likely to be alpha enhanced, we expect the real temperature offsets at low metallicity to be slightly ($\sim$15 K) smaller than shown in Figure \ref{Fig:mas2008}. 
We note that this slope is significantly larger than the one measured using the APOKASC data, this suggests that there could be correlated metallicity and temperature errors in this sample. We suspect that this could result from the heterogeneous determination of the stellar parameters. Given that the offset seen is much larger than expected, we find this test inconclusive as well and suggest that it should be repeated as radii and spectroscopy for stars in the recent {\it Gaia} release \citep{Gaia1} become available. For the purposes of this paper, we will assume that the metallicity-dependent temperature offset measured in the APOKASC sample is real, rather than the result of additional unrecognized systematic errors. 

\section{Results}
\subsection{A Metallicity-Dependent Mixing Length}
While there are a variety of model ingredients, we suspect that the
 assumption of a universal solar calibrated convective efficiency may be causing the observed correlation between temperature offset and metallicity. We explore what metallicity dependent changes to the mixing length $\alpha$ would need to be made to bring the models into agreement with the observations. Specifically, we calculate an ``effective mixing length'' for each star by comparing the offset between the measured $T_{\rm eff}$ and the model predicted $T_{\rm eff}$ for three different mixing lengths. We then quadratically interpolate between them to find the mixing length that produces a temperature offset of 0~K. The mixing length for red giants at solar metallicity is not solar, suggesting a temperature or evolutionary state dependence in the mixing length.

The results are shown in Figure \ref{Fig:ml}, and indicate a metallicity-dependent mixing length. We note that the scatter in effective mixing length (0.065) is about half the uncertainty we would expect given the quoted temperature uncertainties alone (69 K, $\Delta\alpha\sim$0.1), indicating that there is very little actual scatter in our metallicity-effective mixing length relation. We suggested in Section 3.1 that there were possible nonlinearities in the fit of the temperature offset with metallicity which also show up in the mixing length fit. Except for these possible nonlinearities, there do not seem to be strong residual correlations between the mixing length and the $\alpha$-element enhancement, mass, gravity, or temperature although we can not conclusively rule out smaller correlations at this point. We therefore report that a mixing length of $ \alpha_{ML}= 0.162 {\rm [Fe/H]}+ 1.90 $ would best bring our YREC grid into agreement with the APOKASC data. Preliminary investigations indicate that the change in giant branch temperature with mixing length is also linear and similarly sized in the GARSTEC grid of models \citep{WeissSchlattl2008}. We therefore suggest that while the normalization will be different for different model grids, the slope is unlikely to change appreciably. While the model atmospheres used to measure the temperatures also assume a mixing length, our tests indicate that the model spectra, and therefore the derived parameters, are not very sensitive to the mixing length assumed. A change in the mixing length of the atmosphere model of the magnitude discussed here will therefore not impact our results.

We note that detailed seismic modeling of dwarfs and subgiants suggests a range of mixing lengths that depend on temperature, composition, and gravity \citep[see e.g.][]{Bonaca2012, Metcalfe2014, Creevey2016, SilvaAguirre2017}. However, none of these studies have more than a single star in the more evolved gravity ranges we are considering. We therefore assume that this difference from the strongly metallicity dependent temperature offset that we suggest here tentatively supports the idea that our changes in the effective mixing length are actually correcting for metallicity dependent physics errors unique to giants. However, even if this is the case, it is still possible to  mitigate the offset between the data and the models in order to compute accurate isochrone masses and ages from {\it Gaia} data by calibrating the effective mixing length.

\subsubsection{ Comparison with Convection Simulations}
\begin{figure}
\begin{center}
\subfigure{\includegraphics[width=9cm, clip=true, trim=0cm 0cm 0cm 0.8cm]{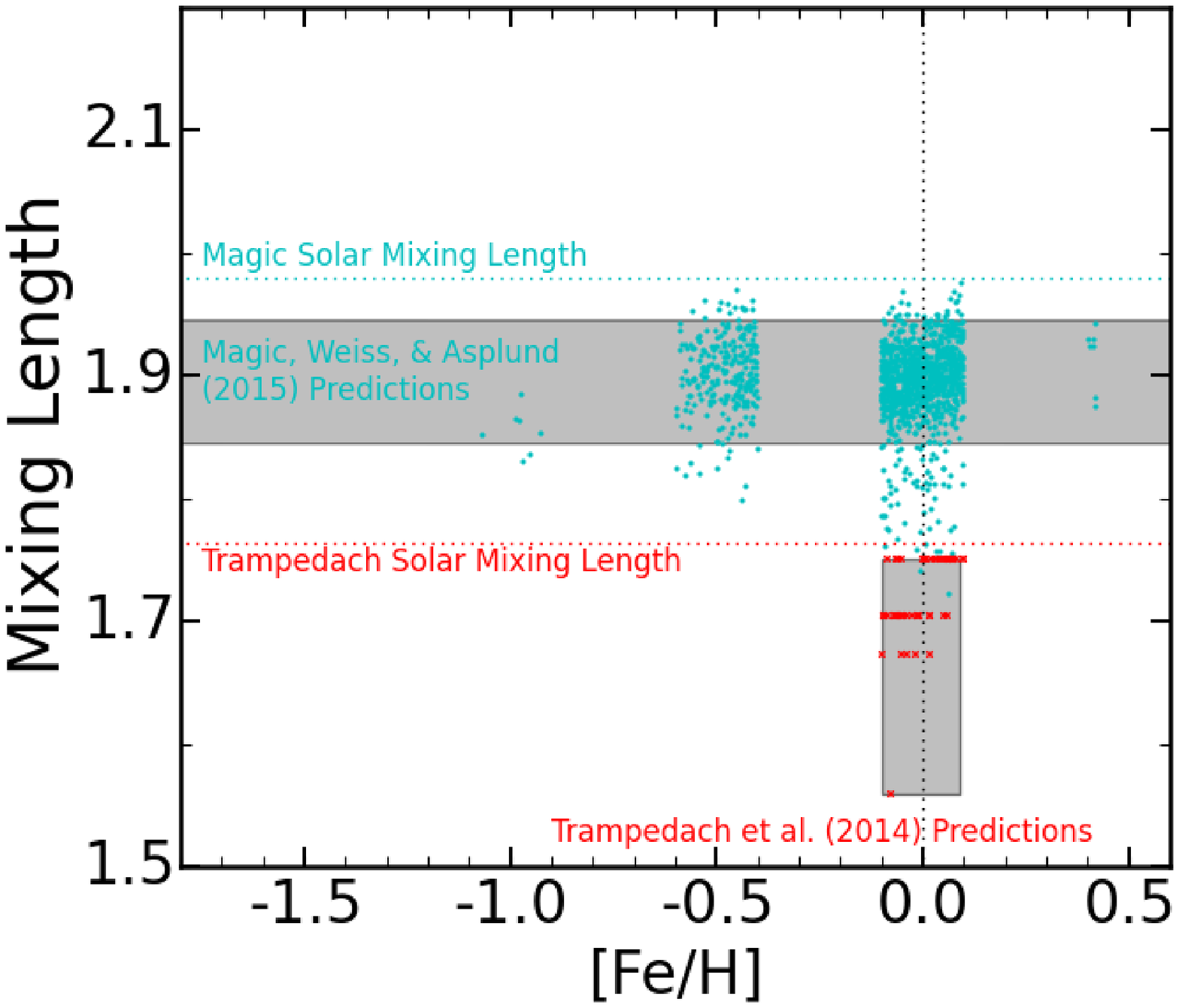}}
\subfigure{\includegraphics[width=9cm, clip=true, trim=0cm 0cm 0cm 0.8cm]{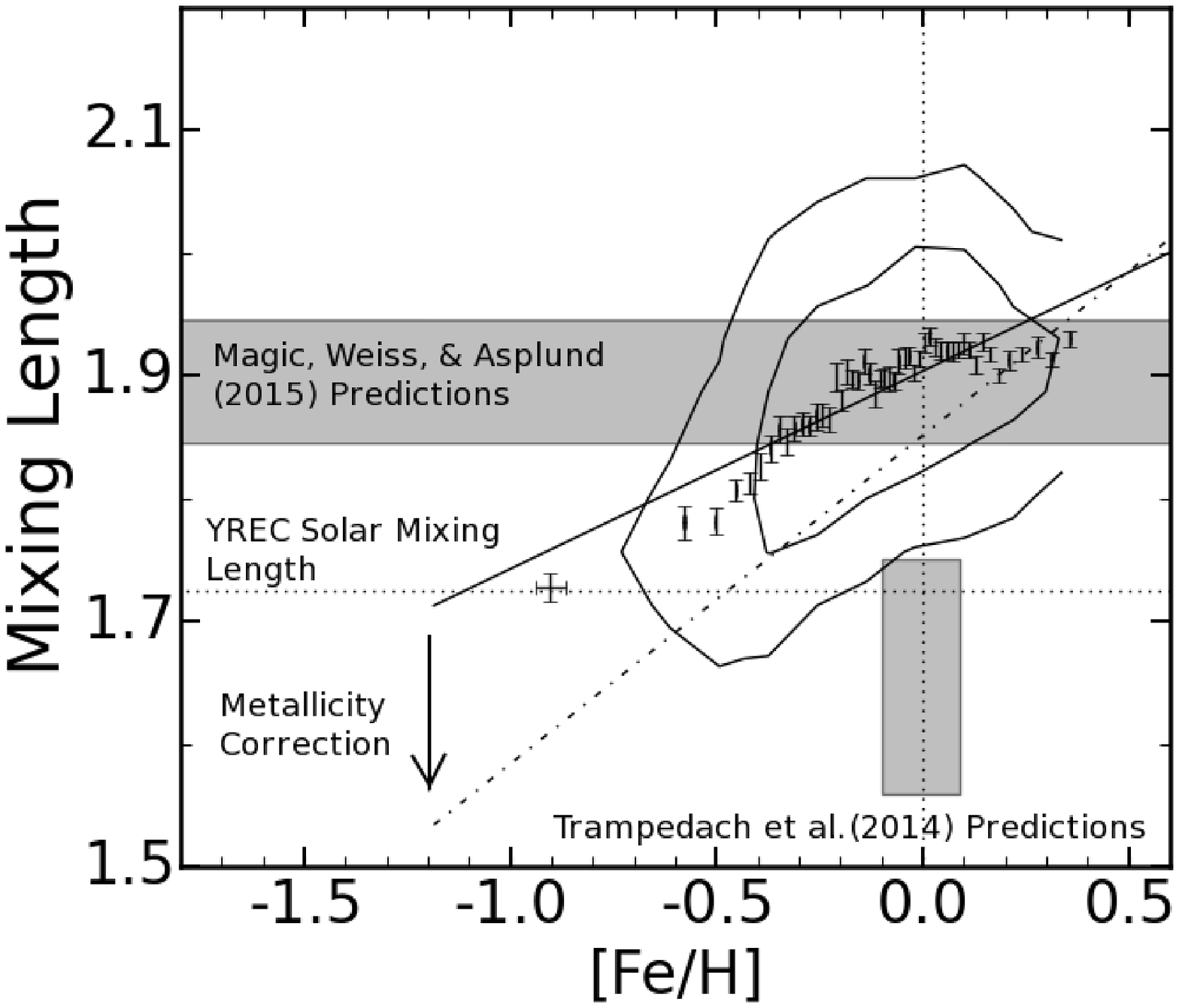}}
\subfigure{\includegraphics[width=9cm, clip=true, trim=0cm 0cm 0cm 0.8cm]{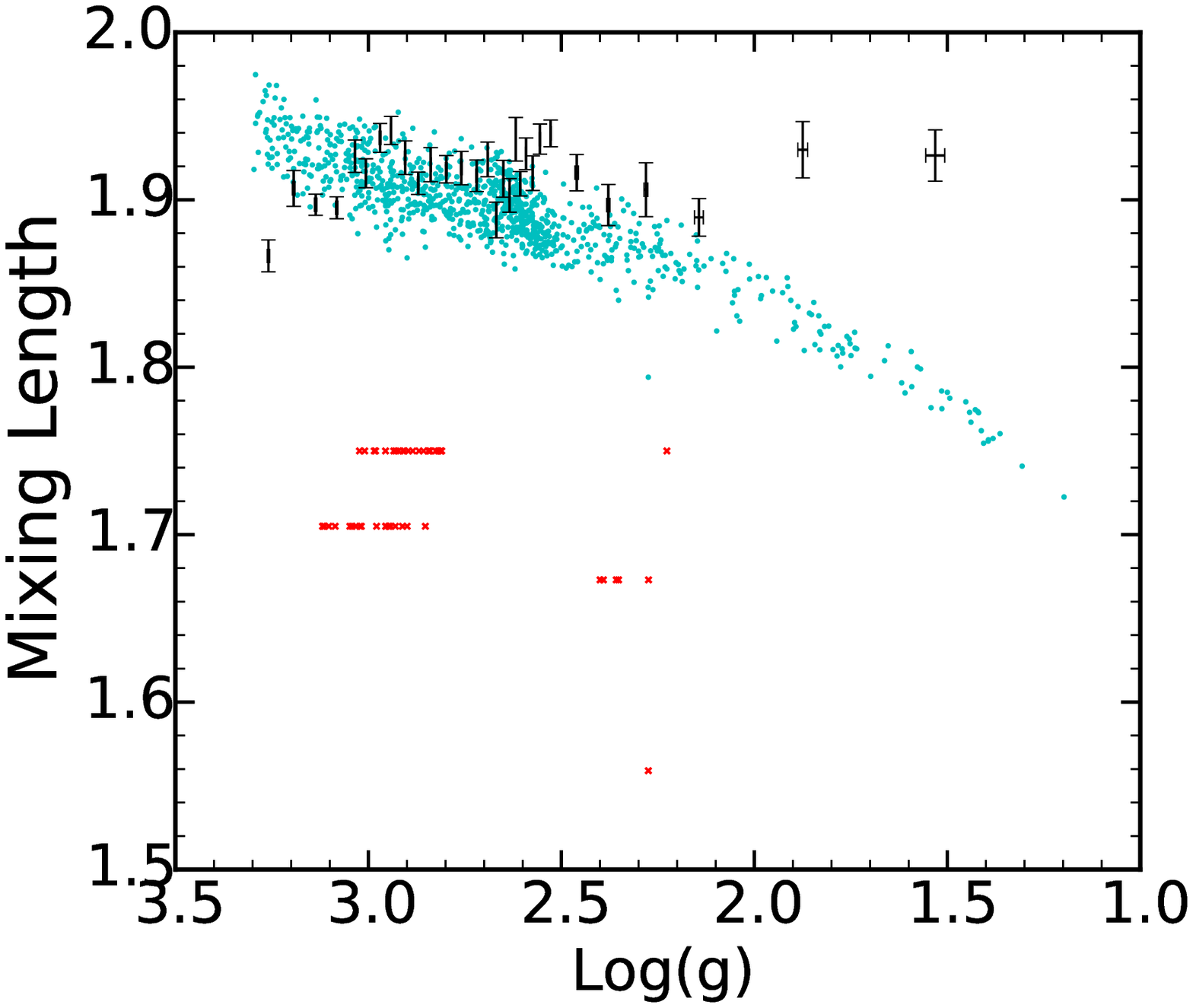}}
\caption{ Top: Mixing length predicted for the temperatures and gravities of the stars in our sample at the metallicites available from the \citet[][$\alpha_{ML, \sun}=1.98$]{Magic2015} (cyan) and \citet[][$\alpha_{ML, \sun}=1.764$]{Trampedach2014} (red) simulations.  Middle: Effective mixing length needed to match the observed data as a function of metallicity for our YREC models ($\alpha_{ML, \sun}=1.72$). We have divided the data into fifty equally sized bins in metallicity, black error bars represent the mean and standard deviation of each bin. The best fit,  $ \alpha_{ML}= 0.1612 {\rm [Fe/H]}+ 1.9037 $, is shown as a solid line; the dot dashed line indicates the best fit using the metallicity corrections discussed in Section 2. Gray bands indicate the range of predicted mixing lengths for our data set from the simulations. Bottom: Mixing length predicted by the simulations (same color coding as top) for stars in our sample with solar metallicity ($-0.1<$[Fe/H]$<0.1$) compared to the effective mixing length required for these stars (black error bars, 30 equally sized bins).  } 
\label{Fig:ml}
\end{center}
\end{figure}

 We also show in Figure \ref{Fig:ml} the range of mixing lengths predicted for the stars in our sample by the Stagger \citep[][$ \alpha_{ML, \sun}=1.98$]{Magic2015} and the \citet[][$ \alpha_{ML, \sun}=1.764$]{Trampedach2014} grids of three dimensional convection simulations. For each grid we use the stars in our sample within 0.1 dex of the metallicites at which the mixing lengths were provided, and we use the temperatures and gravities of those stars to compute a predicted mixing length (top panel). For comparison with the empirical results (middle panel), we mark in gray the full range of results for the Trampedach case. For the Magic sample, we show in gray a mean fit and a two sigma spread. We remind the reader that the \citet{Magic2015} grid uses a substantially different solar mixing length than the YREC grid does, and thus the agreement at solar metallicity might be a coincidence. 

We note that the theoretical predictions from the simulations are inconsistent with the range of effective mixing lengths we fit using one dimensional stellar models. Specifically, to explain the observed temperatures of the stars in our sample would require a much stronger metallicity dependence than the simulations predict (middle panel), at least in some regimes. We also compare the predictions of the simulations as a function of gravity for stars around solar metallicity ($-0.1<$[Fe/H]$<0.1$) to our effective mixing lengths (bottom panel). It is clear that the simulations predict a trend with gravity that is not observed in our sample, and that the spread of mixing lengths at fixed gravity is larger in the simulations than our sample would require to explain the observed effective temperatures. Disagreements with the predicted trend direction and spread have also been observed for dwarf stars whose mixing lengths were obtained from detailed seismic modeling \citep{SilvaAguirre2017}. Together, these discrepancies could indicate either that there is physics not captured or correctly converted to the 1D-mixing length proxy from the 3D-simulations, or that the temperature offset between the data and the models is not entirely due to a change in convection properties on the giant branch, but rather due to other metallicity dependent errors in our stellar evolution model physics.

\subsection{Implications for Age Measurements}

\begin{figure}
\begin{center}
\subfigure{\includegraphics[width=9cm]{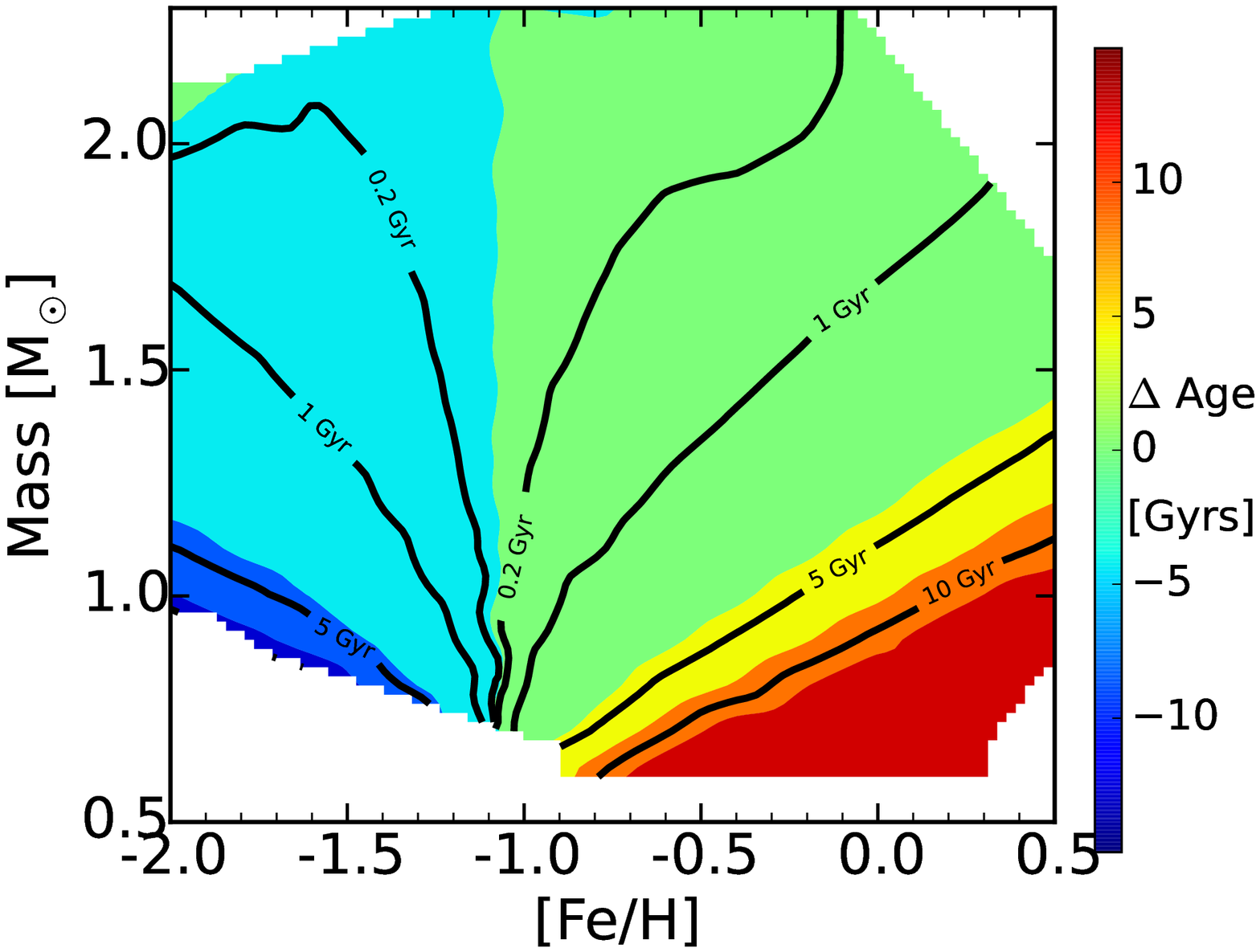}}
\subfigure{\includegraphics[width=9cm]{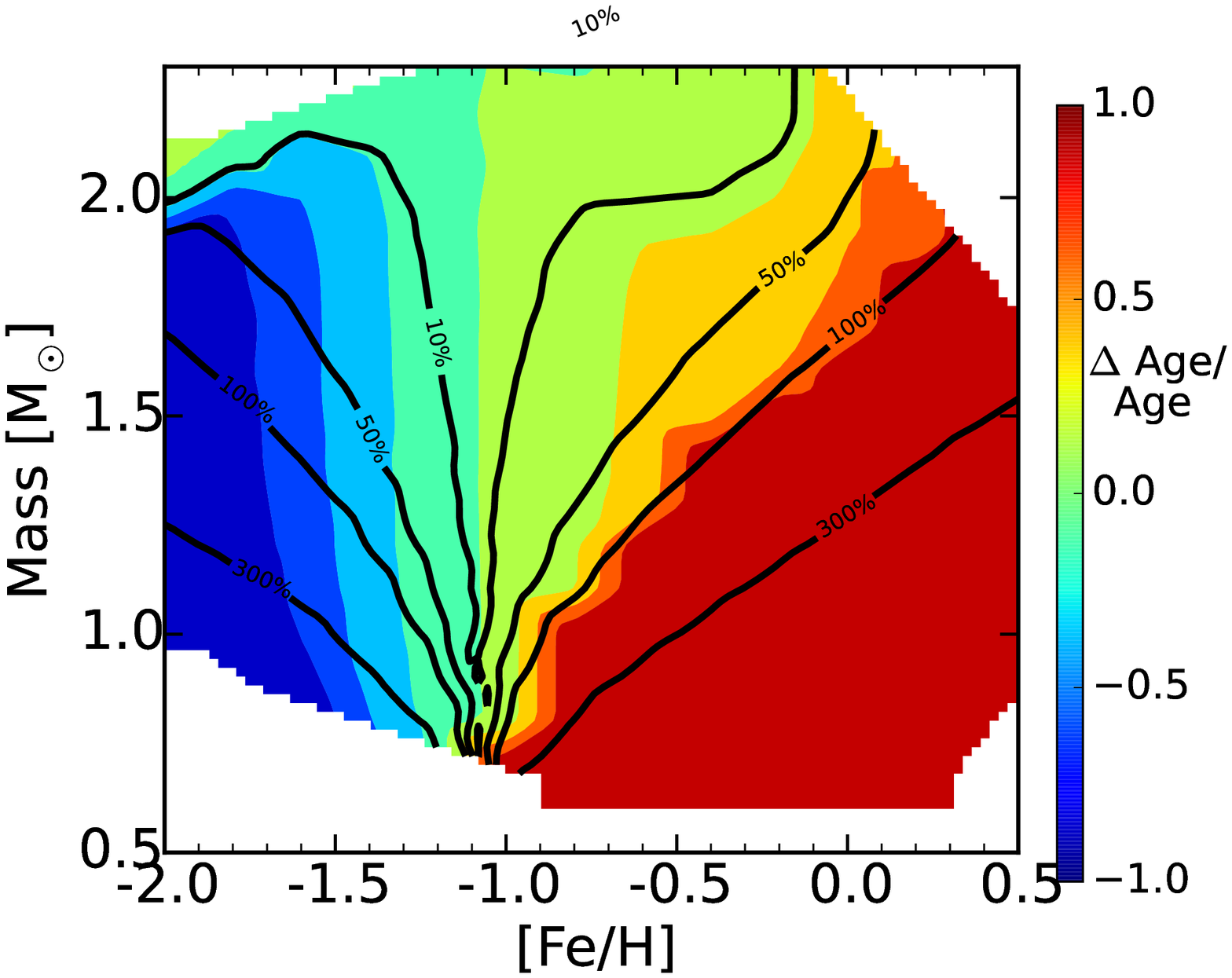}}
\caption{Top: The difference between the real age, assuming our mixing length formula, and the age that would have been inferred from a fixed solar mixing length isochrone of the correct metallicity for a 100 L$_\sun$ star (colors). We have removed points that required extrapolation outside our grid, and excluded stars whose actual age would be greater than 15 Gyrs, because they should not exist in the real \textit{Gaia} sample. Bottom: A similar plot in fractional age, with contours and colors indicating percent and fractional errors respectively. We have truncated the color scheme at age errors of 100\% for clarity. Above about two solar masses, low metallicity stars can still be crossing the Hertzsprung gap as subgiants evolving towards the red giant branch locus at a 100 L$_\sun$, which reduces the impact of the giant branch temperature offset. Because of these large errors, ages should not be computed for \textit{Gaia} stars using solar mixing length isochrones.}
\label{Fig:gaia}
\end{center}
\end{figure}

This correlation between metallicity and temperature offset complicates the interpretation of the recent \textit{Gaia} data. Combining a photometric temperature (and possibly metallicity) with the luminosity inferred from the parallax will not allow for accurate isochrone fitting of masses, and therefore ages, of red giants using standard stellar models. Given our observed temperature offset, we would expect isochrone masses computed from uncorrected models to be off by as much as 0.2 M$_\sun$ ($\sim$ 50 K) at [Fe/H] of $\pm$0.5, even in the optimistic case where the solar metallicity tracks have been properly calibrated. This corresponds to ages incorrect by about a factor of two, which is several gigayears in the solar mass case. Moreover, low metallicity stars will be inferred as less massive (and therefore older) while high metallicity stars will be inferred as more massive (therefore younger), giving a shallower age-metallicity relation in the local galaxy.  We show in Fig~\ref{Fig:gaia} the expected difference between a star's real age and the age that would be inferred using a solar mixing length isochrone for stars whose actual and inferred masses fall within our grid of models and whose actual age is less than 15 Gyr.  We note that the age errors are particularly large for low-mass stars, and can become larger than five gigayears for low-mass, low-metallicity stars. We therefore emphasize the need for isochrones to be made at a variety of mixing lengths, which can be calibrated to correct for these effects. 

\subsection{Implications for Nucleosynthesis}
If the temperature offset is caused by a real change in convection, this would have important implications for the stellar nucleosynthesis models of low- and intermediate-mass stars (1 $<$ M $<$ 10 M$_\sun$). This is because theoretical stellar nucleosynthesis models usually assume the same solar calibrated mixing length parameter for different metallicities and evolutionary stages including the red giant branch and asymptotic giant branch phases \citep[see e.g.,][]{Lugaro2012, KarakasLattanzio2014}. For example, a higher mixing length parameter produces a higher horizontal branch temperature and the predicted yields for massive AGB and super-AGB stars may vary significantly as a result (for some elements, up to a factor of 3; see e.g., \citealt{Doherty2014}).

\subsection{Implications for Stars in the Instability Strip}
We have noted here that changes in the mixing length are required to match giant branch observations. Previous work on subgiant stars has indicated a definite, although possibly smaller, metallicity dependence for the mixing length \citep{Bonaca2012, Metcalfe2014}. We suspect, therefore, that there is likely a metallicity dependence for the mixing length of all stars between the main sequence and the main giant branch locus, including those in the RR Lyrae and Cepheid instability strips. Changing the stellar mixing length affects the minimum masses of stars that develop the Cepheid loop, with the reduced mixing length at low metallicity making it harder to reach the instability strip and hence increasing the minimum Cepheid mass. Similarly, the reduced mixing length at low metallicity will make it hard for low-mass stars to reach the RR Lyrae instability strip and hence will require even more mass loss than commonly invoked to explain the CMD of low-metallicity globular clusters.

\subsection{Implications for Galaxies}
The physical properties of extra-galactic unresolved stellar populations like star clusters and galaxies are obtained by comparing observational spectrophotometry with so-called evolutionary population synthesis models \citep[e.g.][etc.]{Maraston1998, Maraston2005, BruzualCharlot2003, Conroy2009}. These models, which provide integrated spectral energy distributions and mass-to-light ratios as a function of time, metallicity, initial mass function, etc., are based on stellar evolutionary tracks. Hence they are affected by the input physics and the unknowns of stellar evolution. Relevant to this work, \citet{Maraston2005} describes how the mixing length adopted in different evolutionary calculations affect the temperature-luminosity relation of the red giant branch (their Figure 9). Because the red giant branch contributes a large part of the bolometric energy \citep[$\sim$~30-40\%,][] {Maraston1998, Maraston2005} of old  stellar populations ($t > $~a few Gyr), the adopted mixing length impacts population synthesis models. For example, the theoretical $V-K$~color of the models can be affected by $\sim0.25$~mag (their Figure 27). When population synthesis models are fit to data, the derived physical properties will depend on the model input physics. In Goddard et al. (2016, in prep), we find that MaNGA galaxies that are fit with a warmer red giant branch are older by $\sim$~2.5 Gyr around absolute ages of 10 Gyr than galaxies fit to a cooler giant branch. Given the possible offset from solar mixing length, the possible nonlinearity of the mixing length we propose here, and the correlation between metallicity and galaxy mass, we suggest that more work needs to be done to understand the impact of our temperature offset trend on our understanding of galaxies. Galaxy ages and metallicities are key to understanding galaxy formation and evolution in the Universe, hence it is important to realize that the calibration of the mixing length carries cosmological implications. 

\section{Why Wasn't This Noticed Before?}

Given the strength of this effect, it seems surprising that it was not previously recognized. In fact, offsets from isochrones have been noticed both in metal-poor globular clusters \citep[e.g.][]{Brasseur2010,Cohen2015} and metal-rich bulge stars \citep{Ness2013}. Additionally, work on open clusters has indicated that depending on the choice of models, isochrones can be too blue in some bands and too red in others \citep[see e.g.][]{Hayes2015}. However, since each of these regimes is usually analyzed individually, using each author's choice of models and colors, this likely made it more difficult to identify a systematic offset between the models and the data. We suspect that the calibration of the color-temperature relations and the reddening assumptions could also be masking the offset. In low metallicity clusters, for example, where the model giant branch was bluer than the data, it is likely that increased reddening or higher metallicity were assumed during the isochrone fitting in order to make the models better match the data. When combined with the fact that most authors studying star clusters do qualitative, rather than quantitative, assessments of the accuracy of their giant branch locations after fitting to the location of the main sequence or main sequence turnoff, it is somewhat less surprising that these offsets of only about 100 K were not noticed previously. 

 We also suggest that the lack of attention paid to these offsets could be due partially to the fact that the effect is much more obvious in the theoretical gravity-temperature plane than in the observational plane of color and magnitude. One can see this in, for example, the recent work by \citet{Smiljanic2016} on the metal-rich cluster Trumpler 20. Their Figure 2 shows an isochrone which is reasonably close to the data in a color-magnitude diagram, but it is clear in their Figure 3 that the same isochrone is over a hundred Kelvin too cool at fixed gravity.  

\section{Summary}

When comparing the APOKASC data to grids of stellar models, we see a clear, metallicity-dependent offset in temperature. While the exact offset depends on the calibration of the data and the stellar models used, the trend with metallicity is consistent with $\Delta T_{\rm eff} \gtrsim$ 100 K dex$^{-1}$ for the models and calibrations we tested, with $\Delta T_{\rm eff, YREC}= 93.1 {\rm [Fe/H]}+ 107.5$ K for the YREC grid of models. We find no residual dependence of the offset with mass, surface gravity, temperature, or [$\alpha$/Fe] ratio. The offset is difficult to study in optical spectra and cluster data and we suggest that more work should be done in both regimes to determine whether this effect is in fact real and present in all data sets. The models can be brought into agreement with the data if a metallicity dependent mixing length is used, with the required correction for the YREC models being $ \alpha_{\rm ML, YREC}= 0.161 {\rm [Fe/H]}+ 1.90 $. We note that this predicts a non-solar mixing length for solar metallicity giants and does not agree with the temperature, gravity, and metallicity trends predicted by three dimensional convection simulations. We assert that quantification of this metallicity dependent temperature offset is particularly important for the calculation of ages from the recent {\it Gaia} data, as uncorrected isochrones give ages that are wrong by about a factor of two, even at modest (0.5 dex) deviations from solar metallicity.

\begin{acknowledgements}
We would like to thank J. Choi, C. Conroy, and A. Dotter for helpful discussions which led to discoveries of systematic errors in the APOGEE DR13 temperatures. We would also like to thank J. Zinn, M. Ness, and C. Hayes for pointing out useful references. Finally, we thank the referee for reports that improved the clarity and quality of the manuscript.
 
Funding for the Sloan Digital Sky Survey IV has been provided by
the Alfred P. Sloan Foundation, the U.S. Department of Energy Office of
Science, and the Participating Institutions. SDSS-IV acknowledges
support and resources from the Center for High-Performance Computing at
the University of Utah. The SDSS web site is www.sdss.org.

SDSS-IV is managed by the Astrophysical Research Consortium for the 
Participating Institutions of the SDSS Collaboration including the 
Brazilian Participation Group, the Carnegie Institution for Science, 
Carnegie Mellon University, the Chilean Participation Group, the French Participation Group, Harvard-Smithsonian Center for Astrophysics, 
Instituto de Astrof\'isica de Canarias, The Johns Hopkins University, 
Kavli Institute for the Physics and Mathematics of the Universe (IPMU) / 
University of Tokyo, Lawrence Berkeley National Laboratory, 
Leibniz Institut f\"ur Astrophysik Potsdam (AIP),  
Max-Planck-Institut f\"ur Astronomie (MPIA Heidelberg), 
Max-Planck-Institut f\"ur Astrophysik (MPA Garching), 
Max-Planck-Institut f\"ur Extraterrestrische Physik (MPE), 
National Astronomical Observatory of China, New Mexico State University, 
New York University, University of Notre Dame, 
Observat\'ario Nacional / MCTI, The Ohio State University, 
Pennsylvania State University, Shanghai Astronomical Observatory, 
United Kingdom Participation Group,
Universidad Nacional Aut\'onoma de M\'exico, University of Arizona, 
University of Colorado Boulder, University of Oxford, University of Portsmouth, 
University of Utah, University of Virginia, University of Washington, University of Wisconsin, 
Vanderbilt University, and Yale University.

JT, MHP, and JAJ acknowledge support from NASA grant NNX15AF13G. JAJ also acknowledges support from NSF grant AST-1211673. The research leading to the presented results has received funding from the European Research Council under the European Community's Seventh Framework Programme (FP7/2007-2013) / ERC grant agreement no 338251 (StellarAges). GS and JCB acknowledge the support of the Vanderbilt Office of the Provost through the Vanderbilt Initiative in Data-intensive Astrophysics (VIDA). Szabolcs M{\'e}sz{\'a}ros has been supported by the Premium Postdoctoral
Research Program of the Hungarian Academy of Sciences, and by the Hungarian
NKFI Grants K-119517 of the Hungarian National Research, Development and Innovation Office. DAGH was funded by the Ram{\'o}n y Cajal fellowship number RYC-2013-14182. DAGH and OZ acknowledge
support provided by the Spanish Ministry of Economy and Competitiveness (MINECO) under grant AYA-2014-58082-P. FB is supported by NASA through Hubble Fellowship grant \#HST-HF2-51335 awarded by the Space Telescope Science Institute, which is operated by the Association of Universities for Research in Astronomy, Inc., for NASA, under contract NAS5-26555. RAG and BM acknowledge the support form the CNES. SB acknowledged support from NSF grant AST-1514676 and NASA grant NNX16AI09G. Funding for Kepler Discovery Mission is provided by NASA's Science Mission Directorate. Funding for the Stellar Astrophysics Centre is provided by The Danish National Research Foundation (Grant DNRF106). VSA acknowledges support from VILLUM FONDEN (research grant 10118). REC acknowledges funding through Gemini-CONICYT Project 32140007. AS acknowledges support from grants 2014-SGR-1458 (Generalitat de Catalunya), ESP2014-56003-R and ESP2015-66134-R (MINECO). D.H. acknowledges support by the Australian Research Council's Discovery Projects funding scheme (project number DE140101364) and support by the National Aeronautics and Space Administration under Grant NNX14AB92G issued through the Kepler Participating Scientist Program. SM would like to acknowledge support from NASA grants NNX12AE17G and NNX15AF13G and NSF grant AST-1411685.

\end{acknowledgements}

\bibliographystyle{apj} 
\bibliography{ms}
\end{document}